\documentclass[%
reprint,
superscriptaddress,
amsmath,
amssymb,
aps,
prx,
floatfix,
]{revtex4-2}

\usepackage{graphicx}
\usepackage{dcolumn}
\usepackage{bm}
\usepackage{hyperref}
\usepackage{xr}
\usepackage{algorithm}
\usepackage[end]{algpseudocode}

\begin{document}
\title{Phase-dependent microwave response of a graphene Josephson junction}
\author{R.~Haller}
 \email{roy.haller@unibas.ch}
\affiliation{
Department of Physics, University of Basel, Klingelbergstrasse 82 CH-4056, Switzerland
}
\author{G.~F{\"u}l{\"o}p}
\affiliation{
Department of Physics, University of Basel, Klingelbergstrasse 82 CH-4056, Switzerland
}
\affiliation{%
Department of Physics, Budapest University of Technology and Economics and MTA-BME “Momentum” Nanoelectronics Research Group, H-1111 Budapest, Budafoki {\'u}t 8., Hungary
}
\author{D.~Indolese}
\affiliation{
Department of Physics, University of Basel, Klingelbergstrasse 82 CH-4056, Switzerland
}
\author{J.~Ridderbos}
\affiliation{
Department of Physics, University of Basel, Klingelbergstrasse 82 CH-4056, Switzerland
}
\author{R.~Kraft}
\affiliation{
Institute of Nanotechnology, Karlsruhe Institute of Technology, D-76021 Karlsruhe, Germany
}
\affiliation{
Institute of Physics, Karlsruhe Institute of Technology, D-76131 Karlsruhe, Germany
}
\author{L.\,Y.~Cheung}
\affiliation{
Department of Physics, University of Basel, Klingelbergstrasse 82 CH-4056, Switzerland
}
\author{J.\,H.~Ungerer}
\affiliation{
Department of Physics, University of Basel, Klingelbergstrasse 82 CH-4056, Switzerland
}
\affiliation{
Swiss Nanoscience Institute, University of Basel, Klingelbergstrasse 82 CH-4056, Switzerland
}
\author{K.~Watanabe}
\affiliation{
Research Center for Functional Materials,\\ National Institute for Materials Science, 1-1 Namiki, Tsukuba 305-0044, Japan
}
\author{T.~Taniguchi}
\affiliation{
International Center for Materials Nanoarchitectonics, National Institute for Materials Science, 1-1 Namiki, Tsukuba 305-0044, Japan
}
\author{D.~Beckmann}
\affiliation{
Institute of Quantum Materials and Technologies, Karlsruhe Institute of Technology, D-76021 Karlsruhe, Germany
}
\author{R.~Danneau}
\affiliation{
Institute of Quantum Materials and Technologies, Karlsruhe Institute of Technology, D-76021 Karlsruhe, Germany
}
\author{P.~Virtanen}
\affiliation{
Department of Physics and Nanoscience Center, University of Jyv{\"a}skyl{\"a}, P.O. Box 35 (YFL), University of Jyv{\"a}skyl{\"a} FI-40014, Finland
}
\author{C.~Sch{\"o}nenberger}
\homepage{http://www.nanoelectronics.unibas.ch/}
\affiliation{
Department of Physics, University of Basel, Klingelbergstrasse 82 CH-4056, Switzerland
}
\affiliation{
Swiss Nanoscience Institute, University of Basel, Klingelbergstrasse 82 CH-4056, Switzerland
}

\date{\today}

\begin{abstract}

Gate-tunable Josephson junctions embedded in a microwave environment provide a promising platform to in-situ engineer and optimize novel superconducting quantum circuits.
The key quantity for the circuit design is the phase-dependent complex admittance of the junction, which can be probed by sensing an rf SQUID with a tank circuit.
Here, we investigate a graphene-based Josephson junction as a prototype gate-tunable element enclosed in a SQUID loop that is inductively coupled to a superconducting resonator operating at 3\,GHz. 
With a concise circuit model that describes the dispersive and dissipative response of the coupled system, we extract the phase-dependent junction admittance corrected for self-screening of the SQUID loop.
We decompose the admittance into the current-phase relation and the phase-dependent loss and as these quantities are dictated by the spectrum and population dynamics of the supercurrent-carrying  Andreev bound states, we gain insight to the underlying microscopic transport mechanisms in the junction.
We theoretically reproduce the experimental results by considering a short, diffusive junction model that takes into account the interaction between the Andreev spectrum and the electromagnetic environment, from which we deduce a lifetime of $\sim17$\,ps for non-equilibrium populations.

\end{abstract}
\maketitle


\section{\label{sec:intro}Introduction}
For Josephson junctions (JJs), in which the superconducting electrodes are linked with a short normal-conducting region, the coherent superconducting interaction is promoted by so-called Andreev bound states (ABSs)~\cite{Kulik1969}.
The material and geometrical properties of the weak link together with the superconducting phase difference $\varphi$ across the JJ define the energy of the ABSs~\cite{Golubov2004}.
Their structure and occupation dynamics determine the inductive and dissipative microwave response, i.e. the admittance of the JJ~\cite{Virtanen2011,Kos2013}.
In particular, the inductive response relates to the time-averaged dispersion of the populated ABSs and reflects the phase dependence of the supercurrent $I_s(\varphi)$ across the junction ~\cite{Bagwell1992, Paila2009}, which is known as the current-phase relation (CPR). 
On the other hand, the dissipative response relates to the fluctuations in the ABS population resulting in temporal changes of the supercurrent \cite{Averin1996a,Martin-Rodero1996}. 
The microscopic source for those dynamics are thermally activated or microwave induced short-lived ABS excitations~\cite{Dassonneville2018a}.
Conclusively, the junction admittance, which is the key quantity to engineer high-frequency Josephson circuits, is highly dependent on the underlying microscopic processes.

The junction admittance can be probed as a function of phase by embedding a JJ in an rf SQUID that couples to a resonator~\cite{Chiodi2011a,Ferrier2013,Dassonneville2013ab, Dassonneville2018a, Dou2020}.
The rf SQUID acts as a magnetic flux-tunable complex impedance in the circuit that 
shifts and broadens the resonate behavior, from which one can infer the phase-dependent inductive and dissipative response of the junction~\cite{Lake2017}.
The strong demand for in-situ controllable junctions in microwave applications has raised the attention to JJs consisting of gate-tunable weak links~\cite{Aguado2020}.
Here, we determine the full complex admittance of a Josephson weak link made of graphene, which is a two-dimensional (2D) material with a linear band structure and
excellent gating properties.

Although graphene JJs have already demonstrated their compatibility in different superconducting high-frequency circuits, such as bolometers~\cite{Lee2020,Kokkoniemi2020}, transmon qubits~\cite{Kroll2018a, Wang2019} and tunable microwave cavities~\cite{Schmidt2018a}, only few experiments have addressed the determination of their phase-dependent admittance \cite{schmidt2020probing, Dou2020}. 
While Ref.\,\cite{Dou2020} has been focusing on the phase-dependent dissipation of the junction under the influence of external irradiation and Ref.\,\cite{schmidt2020probing} on the inductive behaviour, we here investigate both the inductive and dissipative response simultaneously by studying the inherent photonic phase-dependent interplay between the sensing resonator and the graphene JJ.
We present a classical, comprehensive circuit model to infer the full complex junction admittance from the reflective response of a graphene rf SQUID coupled to a superconducting microwave resonator operating at $\sim3$\,GHz. 
We further translate this to the CPR and the phase-dependent dissipation as a function of gate voltage, under consideration of the self-screening effect that arises due to the finite inductance of the SQUID loop. We describe our observations within the framework of ABSs and find remarkable agreement between the experimental results and the theoretically predicted microwave response of a short, diffusive junction.


\section{\label{sec:device}Device}

\begin{figure}[t!]
\includegraphics{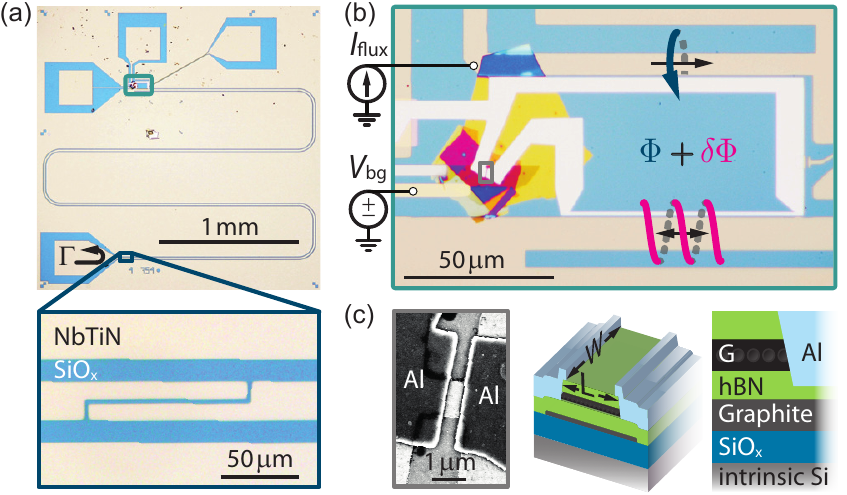}
\caption{\label{fig:device}
Graphene rf SQUID inductively coupled to a superconducting transmission line resonator.
(a) Optical image of the NbTiN $\lambda/4$-resonator consisting of a meandered co-planar transmission line with the shorted end (current anti-node) on top, seen also at the bottom of image 
(b), and the open end (current node) at the bottom, shown in the zoom-in. 
(b) Optical image of the mono\-layer graphene (G) Josephson junction (JJ) embedded in an Al loop forming the rf SQUID. 
The DC current $I_{\rm{flux}}$ creates a flux $\Phi$ inside the loop (blue line), which allows to phase bias the junction. 
The inductive coupling to the resonator induces a small oscillating probe flux $\delta\Phi$ (red lines). 
The gate voltage $V_{\rm{bg}}$ applied on the bottom graphite sheet tunes the charge carrier density in G. 
(c) Scanning electron micrograph and cross-sectional schematics of the hBN-encapsulated G-JJ with Al side-contacts of width ${W=1\,\mu}$m and length ${L=400}$\,nm.
}
\end{figure}

The device is presented in Fig.\,\ref{fig:device} and consists of a graphene JJ embedded in a superconducting loop, which inductively couples to a co-planar transmission line (CTL) resonator.
The resonant structure and supply lines are etched into NbTiN (80\,nm) sputtered on an intrinsic Si/SiO$_x$ (500\,$\mu$m/170\,nm) substrate.
The meandered CTL shown in Fig.\,\ref{fig:device}(a) is shorted to ground on one side, and interrupted by a coupling capacitor on the other.
Both of these terminations act as microwave mirrors of the opposite type, and thereby form a superconducting $\lambda/4$-resonator with a fundamental bare resonance frequency $f_{\rm{bare}}=3.098$\,GHz. 

The graphene JJ, shown in Fig.\,\ref{fig:device}(c), is made of a van der Waals heterostructure consisting of a mono\-layer graphene encapsulated in hexagonal boron nitride (hBN). 
The lower hBN layer (47.5\,nm) separates the graphene flake from the bottom graphite gate.
A thermally evaporated Ti/Al (5\,nm/90\,nm) lead contacts the graphene from both sides \cite{Wang2013} and encloses the junction in a loop, thus forming a graphene rf SQUID, which is inductively coupled to the current anti-node of the resonator as illustrated in Fig.\,\ref{fig:device}(b).
The galvanic grounding of the loop defines the reference potential for the gate voltage $V_{\rm{bg}}$ applied on the bottom graphite structure.
The DC current $I_{\rm{flux}}$ controls the magnetic flux $\Phi$ inside the loop and therefore tunes the external phase difference $\varphi_{\rm{ext}}=2\pi\Phi/\Phi_0$ across the rf SQUID, where $\Phi_0=h/2e$ is the superconducting flux quantum with $h$ being the Planck constant and $e$ the elementary charge.
Consider the Supplemental Material (SM) for details about the device fabrication~\footnote{See Supplemental Material for details about the device fabrication (Sec.\,\ref{sec:fabrication}), measurement scheme and calibrations (Sec.\,\ref{sec:setup},\,\ref{sec:readout} and \ref{sec:chargedensity}), procedures for fitting the resonance curve (Sec.\,\ref{sec:rescurvefit}), derivations and validity proof for the formulas relating the resonant behavior to the electrical properties of the JJ (Sec.\,\ref{sec:loadedres}-\ref{subsec:test_ana_formula}), iterative fitting routine to correct for screening (Sec.\,\ref{sec:screening_correction}) and theoretical as well as experimental results as a function of temperature (Sec.\,\ref{sec:temp_dep})}.

In the subsequent experiment we perform reflectance measurements on the port denoted by $\Gamma$ in Fig.\,\ref{fig:device}(a) and investigate the resonant circuit as a function of $V_{\rm{bg}}$ and $I_{\rm{flux}}$, from which we later infer the CPR and the phase-dependent loss of the graphene JJ.

\section{\label{sec:reflec}Reflectomerty}

The coupled microwave circuit is probed by reflectometry in a dry dilution refrigerator, in which the device is surrounded by a permalloy shield.
With a vector network analyzer we measure the complex reflection coefficient $\Gamma$ as a function of probe frequency $f$ and $I_{\rm{flux}}$.
We ensure a quasi-equilibrium sensing by setting the probe power to an averaged intra-cavity occupation of $\sim100$ photons, which corresponds to an oscillating probe flux $\delta\Phi\approx\Phi_0/100$ inside the SQUID loop.
Additionally, we tune the charge carrier density in the graphene layer by applying a gate voltage in the range $V_{\rm{bg}}=\left[-9,9\right]$\,V.
The conversion from $V_{\rm{bg}}$ to charge carrier density as well as the measurement scheme and the calibration of the probe power can be found in the SM \cite{Note1}.

\begin{figure}[b!]
\includegraphics{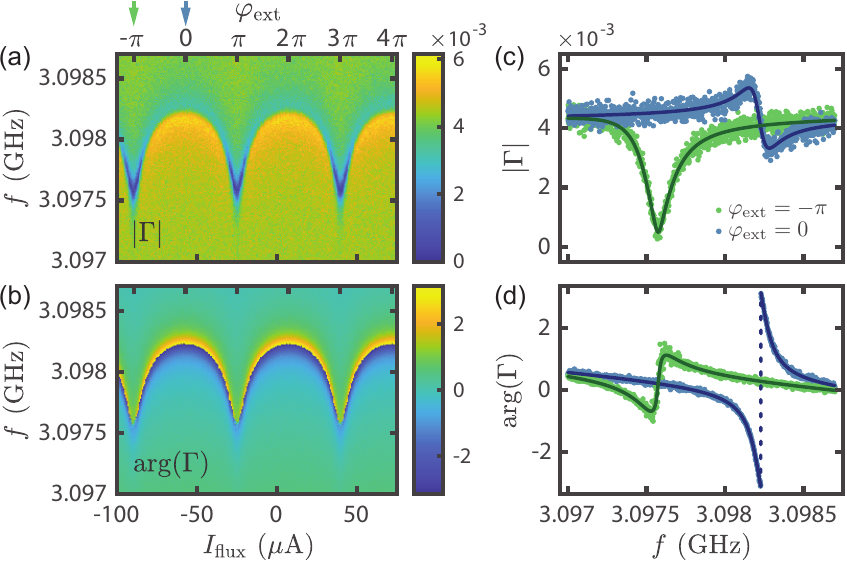}
\caption{\label{fig:res}
Flux dependence of the reflection coefficient $\Gamma$ at ${V_{\rm{bg}}=6}$\,V.
(a)-(b) Colormaps of $\lvert\Gamma\rvert$ and $\rm{arg}\left(\Gamma\right)$ as a function of probe frequency $f$ and DC flux current $I_{\rm{flux}}$.
The horizontal top axis represents the conversion to the external phase $\varphi_{\rm{ext}}$ across the rf SQUID.
(c)-(d) $\lvert\Gamma\rvert$ and $\rm{arg}\left(\Gamma\right)$ at $\varphi_{\rm{ext}}=[-\pi,0]$ overlaid with fits to Eq.~\ref{eq:complex_fit} (solid lines), from which we obtain the resonance frequency
$f_0$, asymmetry angle $\phi$, coupling quality factor $Q_{\rm{c}}$ and effective quality factor $Q_{\rm{e}}$ as listed below:
}
\begin{ruledtabular}
\begin{tabular}{ccccc}
\textrm{$\varphi_{\rm{ext}}$}&
\textrm{$f_0$}&
\textrm{$\phi$}&
\textrm{$Q_c$}&
\textrm{$Q_e$}\\
\colrule
$-\pi$ & $3.09755$\,GHz & $0.224$ & $23\,400$ & $19\,400$\\
$0$ & $3.09821$\,GHz & $0.235$ & $23\,700$ & $>200\,000$\\
\end{tabular}
\end{ruledtabular}
\end{figure}

The reflective response at ${V_{\rm{bg}}=6}$\,V presented in Fig.\,\ref{fig:res} is exemplary for the whole measurement set.
Clear periodic shifts of the resonance frequency $f_0$ as a function of $I_{\rm{flux}}$ can be observed in Fig.\,\ref{fig:res}(a) and Fig.\,\ref{fig:res}(b).
We encounter no phase jumps and relate the external phase $\varphi_{\rm{ext}}=n_{\rm{odd}}\pi$ ($=n_{\rm{even}}\pi$) to points of minimal (maximal) resonance frequencies~\cite{Chiodi2011a,Lake2017}.
Besides $f_0$, the resonance lineshape also changes as seen in Fig.\,\ref{fig:res}(c) and Fig.\,\ref{fig:res}(d) when comparing line cuts at ${\varphi_{\rm{ext}}=-\pi}$ and ${\varphi_{\rm{ext}}=0}$.
As we will show, both the modulation in $f_0$ and the altered lineshape are the consequence of the phase-dependent complex admittance of the graphene JJ.

In order to characterize the JJ from the reflective response, we fit $|\Gamma|$ and $\rm{arg}\left(\Gamma\right)$ simultaneously for each combination of $V_{\rm{bg}}$ and $I_{\rm{flux}}$ with the complex resonance curve of a loaded $\lambda/4$-resonator expressed according to Ref.\,\cite{Khalil2012a} as:
\begin{equation}
\label{eq:complex_fit}
\Gamma= \left[\frac{\Gamma_{\rm{min}}+2j{}Q\frac{f-f_0}{f_0}}{1+2j{}Q\frac{f-f_0}{f_0}}-1\right]e^{j\phi}+1.
\end{equation}

Thus, we can deduce $f_0$ and assess the broadening of the resonance curve.
The latter is determined by the total quality factor $Q=1/(Q_{\rm{load}}^{-1}+Q_i^{-1}+Q_c^{-1})$, which in turn, consists of three different dissipation sources:
i) The inverse load quality factor $Q_{\rm{load}}^{-1}$ describes loss generated by the rf SQUID,
ii) the inverse internal quality factor $Q_i^{-1}$ describes loss inherent to the properties of the CTL and
iii) the inverse coupling quality factor $Q_c^{-1}$ describes loss to the measurement environment.
Here, $Q_{\rm{load}}^{-1}$ and $Q_i^{-1}$ are merged to an effective quality factor ${Q_e=1/(Q_{\rm{load}}^{-1}+Q_i^{-1})}$.
Furthermore, we define ${\Gamma_{\rm{min}}={(Q_c-Q_e)}/{(Q_c+Q_e)}}$ and introduce the angle $\phi$, which accounts for an asymmetric line shape.

The fits to Eq.~\ref{eq:complex_fit} at ${\varphi_{\rm{ext}}=-\pi}$ and ${\varphi_{\rm{ext}}=0}$, shown in Fig.\,\ref{fig:res}(c) and Fig.\,\ref{fig:res}(d) as solid lines, reveal an overall shift of 660\,kHz in $f_0$ and a drastic change in $Q_e$, while $Q_c$ and $\phi$ remain similar.
At $\varphi_{\rm{ext}}=-\pi$, we obtain ${Q_e=19\,400}$ and ${Q_c=23\,400}$; whereas at $\varphi_{\rm{ext}}=0$, we find ${Q_e>200\,000}$ and ${Q_c=23\,700}$.
Consequently, the resonator is undercoupled ($Q_e<Q_c$) at $\varphi_{\rm{ext}}=-\pi$, but overcoupled ($Q_e>Q_c$) at $\varphi_{\rm{ext}}=0$, which explains the distinct resonance lineshapes~\cite{Goppl2008}.
Since $Q_i$ can be treated as a constant with $Q_e$ being a lower bound, we conclude that $Q_i>200\,000$. This large value allows us to treat the CTL as lossless ($Q_i^{-1}=0$) such that ${Q_e\approx Q_{\rm{load}}}$.
The SM provides further insights to the resonance curve fitting~\cite{Note1}.

The observed flux tunable microwave response in terms of $f_0$ and  $Q_{\rm{load}}$ is the direct manifestation of phase-dependent microscopic processes in the graphene JJ~\cite{Chiodi2011a}, which will be discussed in detail in Sec.\,\ref{sec:abs} and Sec.\,\ref{sec:compsimulation} within the framework of ABSs.
In the following section we model the electrical properties of the graphene JJ with lumped elements and explain their effect on the resonant behavior with the circuit of a loaded $\lambda/4$-resonator.


\section{\label{sec:circuitmodel}Circuit model}

\begin{figure}[t!]
\includegraphics{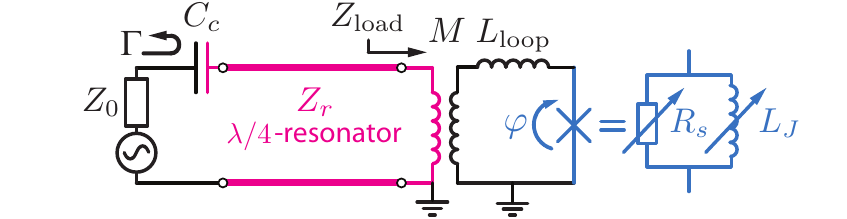}
\caption{\label{fig:circuitmodel} 
Circuit schematic of a rf SQUID coupled to a $\lambda/4$ resonator. 
The resonator couples inductively to the rf SQUID with strength $M$ and connects to the reflectometry setup via capacitance $C_c$. The rf SQUID is modeled as a loop with self-inductance $L_{\rm{loop}}$ in series with the JJ, which in turn, is modeled as a variable Josephson inductance $L_J$ in parallel with a variable shunt resistance $R_s$.
This forms a variable load impedance $Z_{\rm{load}}$, which tunes the reflective response $\Gamma$.
}
\end{figure}

The inductively coupled rf SQUID acts as a variable load impedance $Z_{\rm{load}}$ attached to the resonator, which tunes the reflective response. We express $Z_{\rm{load}}$ according to the circuit schematic depicted in Fig.\,\ref{fig:circuitmodel}.
The rf SQUID is modeled as a loop with self-inductance $L_{\rm{loop}}$ in series with the JJ.
The mutual inductance $M$ quantifies the coupling strength to the resonator, which is built from a CTL with characteristic impedance $Z_r$.
The JJ itself is represented by a variable Josephson inductance $L_J$ in parallel with a variable shunt resistance $R_s$.
For this arrangement the load impedance terminating the resonator is detailed in the SM and reads~\cite{Note1}:
\begin{equation}
\label{eq:zload}
Z_{\rm{load}}=\frac{\omega^2M^2}{j\omega{}L_{\rm{loop}}+\left(G_s+jB_J\right)^{-1}},
\end{equation}
where ${\omega=2\pi{}f}$ is the angular frequency, ${G_s=1/R_s}$ is the shunt conductance and ${B_J=-1/(\omega L_J)}$ is the susceptance.
Note that $Y=G_s+jB_J$ is the complex admittance of the JJ. 

The influence of $Z_{\rm{load}}$ on the $\lambda/4$-resonator is twofold:
First, the imaginary part of $Z_{\rm{load}}$ causes a shift of the resonance frequency as derived in the SM~\cite{Note1}
\begin{equation}
\label{eq:fresfull}
\delta{}f_0=f_0-f_{\rm{bare}}=-\frac{2}{\pi{}Z_r}\operatorname{Im}(Z_{\rm{load}})f_{\rm{bare}},
\end{equation}
with respect to the unloaded resonance frequency $f_{\rm{bare}}$. 
Second, the real part of $Z_{\rm{load}}$ gives rise to dissipation in the resonant circuit, which can be expressed according to the derivations presented in the SM as~\cite{Note1}
\begin{equation}
\label{eq:Qfull}
Q_{\rm{load}}=\frac{\pi{}Z_r}{4\operatorname{Re}(Z_{\rm{load}})}.
\end{equation}

\begin{figure*}
\includegraphics{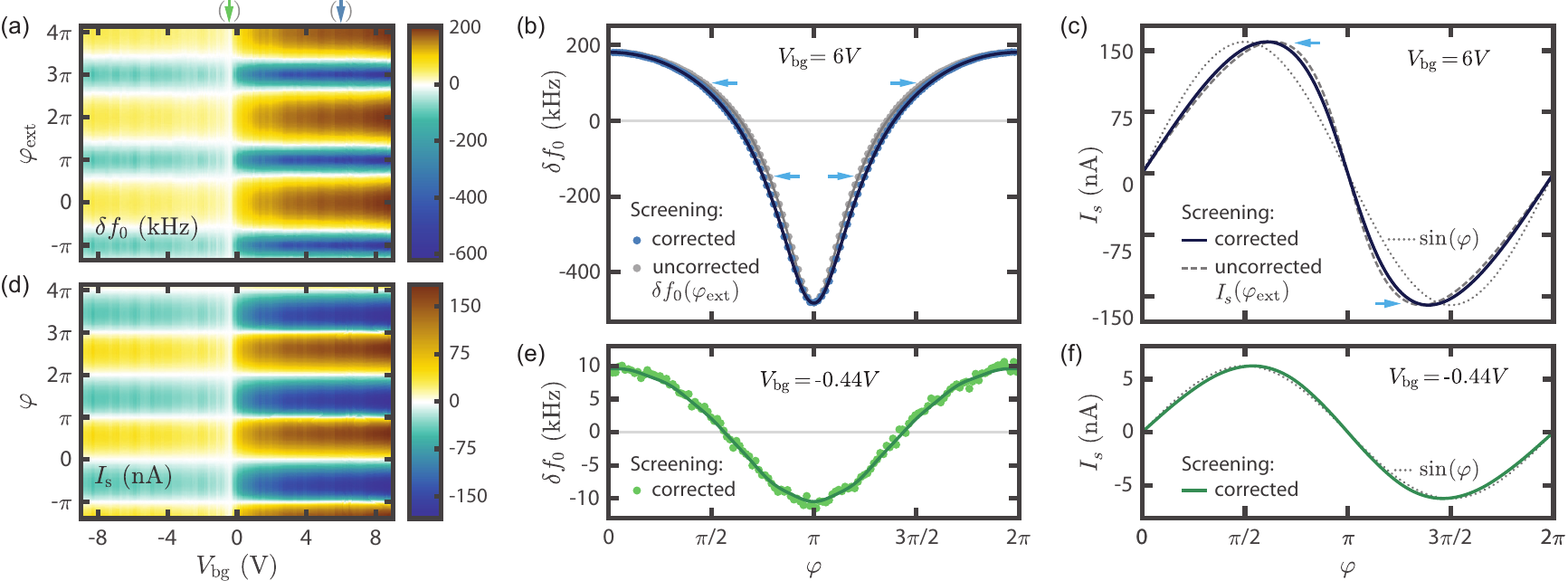}
\caption{\label{fig:cpr}
Evaluation of the CPR.
(a) Colormap of the resonance frequency shift $\delta{}f_0=f_0-f_{\rm{bare}}$ with $f_{\rm{bare}}=3.098$~GHz as a function of gate voltage $V_{\rm{bg}}$ and external phase $\varphi_{\rm{ext}}$.
(b) $\delta{}f_0$ at $V_{\rm{bg}}=6$\,V  as a function of $\varphi$ and $\varphi_{\rm{ext}}$, respectively overlaid with the fits to Eq.~\ref{eq:fres} (solid lines), from which the CPR is deduced.
(c) Presents the CPR at $V_{\rm{bg}}=6$\,V, corrected for the self-screening of the SQUID (blue) and uncorrected (dashed),  in comparison with the sine function (dotted).
In (b)-(c) arrows illustrate the correction introduced by the non-linear mapping from $\varphi_{\rm{ext}}$ to  $\varphi$.
(d) Corrected CPR inferred from (a) as a function of $V_{\rm{bg}}$.
(e) $\delta{}f_0$ at the charge neutrality point ($V_{\rm{bg}}=-0.44$\,V) as a function of $\varphi$ overlaid with the fit and in (f) the corresponding CPR. 
}
\end{figure*}

From Eq.~\ref{eq:zload} one recognizes, that the junction variables, $G_s$ and $B_J$ affect both $\operatorname{Re}(Z_{\rm{load}})$ and $\operatorname{Im}(Z_{\rm{load}})$. 
Consequently, $\delta{}f_0$ and $Q_{\rm{load}}$ would need to be considered simultaneously to evaluate them.
However, it turns out that, due to the obtained relatively large $Q_{\rm{load}}$ values, one is allowed to set $G_s\rightarrow 0$ in Eq.~\ref{eq:fresfull}, which simplifies the relation as shown in the SM to~\cite{Note1} 
\begin{equation} \label{eq:fres}
\delta{}f_0\approx\frac{8}{\pi^2}\frac{M^2}{L_p\left(L_J+L_{\rm{loop}}\right)}f_{\rm{bare}},
\end{equation}
where $L_p$ is the parallel $LC$-equivalent inductance of the $\lambda/4$-resonator.
This means that the shift of the resonance frequency mainly originates from the Josephson inductance $L_J$, whereas the broadening of the resonance originates from the dissipation in the JJ specified by the shunt conductance $G_s$. 

Since the inverse Josephson inductance is a measure of the change in the supercurrent $I_s(\varphi)$ with respect to the phase $\varphi$ across the junction\,\cite{Paila2009} 
\begin{equation}
\label{eq:LCR}
L_J(\varphi)^{-1}=\frac{2\pi}{\Phi_0}\frac{\partial I_s(\varphi)}{\partial\varphi},
\end{equation}
we can express the resonance frequency shift and the behavior of $L_J(\varphi)$ with the current-phase relation (CPR).

In order to quantify the CPR and $G_s$ from the resonator response, we perform finite-element simulations \cite{manual2005sonnet} based on the device geometry, to acquire ${L_{\rm{loop}}=211}$\,pH and $M=30.83$\,pH.
Moreover, we find ${Z_r=69.5\,\Omega}$ from the aspect ratios of the CTL~\cite{Gevorgian1994} in combination with the resonant behavior of the circuit and deduce ${L_p=4.55\,\rm{nH}}$.
The evaluation of $Z_r$ and $L_p$ can be found in the SM~\cite{Note1}.


\section{Current-phase relation}

In this section we extract the CPR by fitting the periodic shift of the resonance frequency under consideration of self-screening effects.
The coupling strength between the superconducting leads is determined by the Cooper pair transmission probability and defines the shape of the CPR.
For small coupling or low transmission probability the CPR is sinusoidal, whereas the CPR becomes forward-skewed for increased coupling. 
Due to the semiconducting properties in graphene JJs, the coupling strength and therefore the CPR skewness can be tuned with the gate voltage~\cite{English2016, Nanda2017, Schmidt2018a, Indolese2020, Manjarres2020}.
To capture the non-sinusoidal behavior, we express the CPR as Fourier series~\cite{Spanton2017}
\begin{equation}
\label{eq:IsFourier}
I_s(\varphi)=\sum_{k}(-1)^{k-1}A_{k}\sin(k\varphi),
\end{equation}
with $k$ being the harmonic order and $A_k$ the corresponding amplitude.

To extract the CPR from the measured resonance frequency modulations, we need to relate the external phase $\varphi_{\rm{ext}}$ to the phase difference $\varphi$ across the JJ.
This is not straightforward, since if a supercurrent flows within the rf SQUID, there is a phase drop over the loop inductance $L_{\rm{loop}}$ in addition to the phase drop over the JJ, which leads to a non-linear relation between the internal phase $\varphi$ and the external phase $\varphi_{\rm{ext}}$ -- known as the screening effect~\cite{Jung2013d}:
\begin{equation}
\label{eq:screening}
\varphi=\varphi_{\rm{ext}}-\frac{2\pi}{\Phi_0}L_{\rm{loop}}I_s(\varphi).
\end{equation}
Here, we obtain the CPR for each gate voltage by solving the set of equations Eqs.\,\ref{eq:fres}-\ref{eq:screening} in a self-consistent way by using an iterative fitting method.
The basis for this method is the resonance frequency shift as a function of $\varphi_{\rm{ext}}$, which is presented for the entire gate range in Fig.\,\ref{fig:cpr}(a).
At each fitting iteration we include Fourier amplitudes $A_k$ up to the $10^{\rm{th}}$-harmonic and allow for small changes in $f_{\rm{bare}}$.
Details about the method can be found in the SM~\cite{Note1}.

In Fig.\,\ref{fig:cpr}(b) we illustrate the effect of screening by comparing $\delta f_0$ as a function of $\varphi$ and $\varphi_{\rm{ext}}$, respectively -- for the example at $V_{\rm{bg}}=6$\,V.
The corresponding CPRs, deduced from fitting the modulations in $\delta f_0$ with respect to phase, shown as solid lines in Fig.\,\ref{fig:cpr}(b), are presented in Fig.\,\ref{fig:cpr}(c).
The screening consideration causes a distortion of the phase around $\pi$ as indicated by arrows.
Omitting this effect results in an apparent enhancement of the skewness~\cite{Nichele2020}.
Even after correcting for screening, we find a substantially forward-skewed CPR, visualized by the comparison with a sinusoidal behavior.
Although screening effects are small in this case, we emphasize that they can have a significant impact on the evaluated skewness, especially for large $I_s$ and $L_{\rm{loop}}$.

In Fig.\,\ref{fig:cpr}(d) we map the extracted CPR as a function of $V_{\rm{bg}}$.
The smallest CPR amplitude is found at $V_{\rm{bg}}=-0.44$\,V, which we attribute to the charge neutrality point (CNP) of graphene.
Here, resonance frequency modulations of only $\pm10$\,kHz can still be clearly resolved as seen in Fig.\,\ref{fig:cpr}(e), which demonstrates the sensitivity of the microwave circuit.
The CPR at the CNP, shown in Fig.\,\ref{fig:cpr}(f), is slightly skewed and has a maximal supercurrent of $I_c=6.3$\,nA.

In the following, we quantify the CPR and its skewness by two commonly used ways:
i) by the skewness parameter $S=(2\varphi_{\rm{max}}/\pi)-1$, where $\varphi_{\rm{max}}$ 
is the phase maximizing the CPR to the critical current $I_c$~\cite{Nanda2017}, and
ii) by directly providing the set of Fourier amplitudes $A_k$~\cite{Spanton2017}.
The latter description is more precise, since it captures the entire CPR lineshape,
whereas the $S$-parameter together with $I_c$ do not uniquely characterize the CPR, but are more intuitive.

\begin{figure}[t!]
\includegraphics{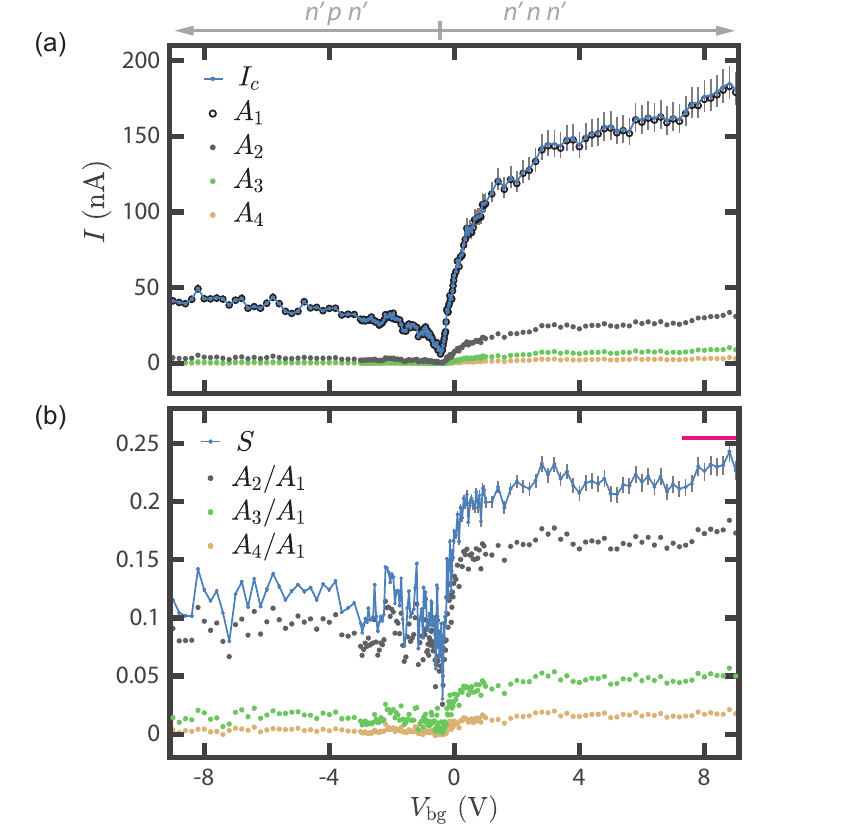}
\caption{\label{fig:cpr2}
Characteristics of the CPR as function of gate voltage $V_{\rm{bg}}$. The step size in $V_{\rm{bg}}$ is reduced close to the CNP ($V_{\rm{bg}}=-0.44$\,V).
(a) Critical current $I_c$ and Fourier amplitudes $A_{\rm{k}}$.
(b) Skewness parameter $S$ and ratios $A_{\rm{k}}/A_1$.
The theoretical skewness value for a short, diffusive system under ideal conditions $S=0.255$ is illustrated with the pink mark.
(a)-(b) Systematic error bars in $I_c$ and $S$ are generated by modifying $M$ by $\pm3\%$ and $L_{\rm{loop}}$ by $\pm5\%$ in the CPR evaluation. The amplitudes $A_k$ for $k\geq5$ are negligibly small and omitted in the figures.}
\end{figure}

In Fig.\,\ref{fig:cpr2} we employ both of these characterizations to illustrate the gate dependence of the CPR.
We observe a rapid enhancement of $I_c$ up to $\sim200$\,nA for gating towards positive voltages ($n$-doped), whereas towards negative voltages ($p$-doped) the increase is weaker and reaches only $\sim50$~nA as seen in Fig.\,\ref{fig:cpr2}(a).
Because $A_1$ closely follows $I_c$, the CPR is mainly determined by the $2\pi$-periodic sinusoidal contribution for all $V_{\rm{bg}}$.
However, the small additions from higher harmonics lead to a forward-skewed CPR.
From Fig.\,\ref{fig:cpr2}(b) it appears that the skewness saturates in both doping regimes with a slight reduction around the CNP.
For the $n$-doped side, the skewness saturates around ${S\approx0.22}$, whereas on the $p$-doped side the skewness is less pronounced, saturating around ${S\approx0.12}$. The ratios $A_k/A_1$ follow the same trend.

The asymmetric behavior in $I_c$ and $S$ with respect to $V_{\rm{bg}}$ are attributed to the presence of $n^\prime$-doped contact regions inducing additional scattering potentials. 
The JJ is therefore more transparent in the $n^\prime{}n\,n^\prime $-situation compared to the $n^\prime{}p\,n^\prime$-case~\cite{Borzenets2016,Nanda2017}.
We speculate that the minimal skewness of $S\approx 0.05$ close to the CNP originates from the formation of electron-hole puddles~\cite{Xue2011} in the graphene flake, which further enhance the scattering probability.


\section{\label{sec:loss}Phase-dependent loss}

Having extracted the CPR from the resonance frequency shift, we now deduce the phase-dependent dissipative part of the graphene JJ; namely, the shunt conductance $G_s$.
We can infer $G_s$ from Eq.\,\ref{eq:Qfull}, in which we express the susceptance $B_J$ with the CPR according to Eq.\,\ref{eq:LCR} and make use of $Q_{\rm{load}}$ obtained from the reflectance curve analysis presented in Sec.\,\ref{sec:reflec}.

\begin{figure}[h!]
\includegraphics{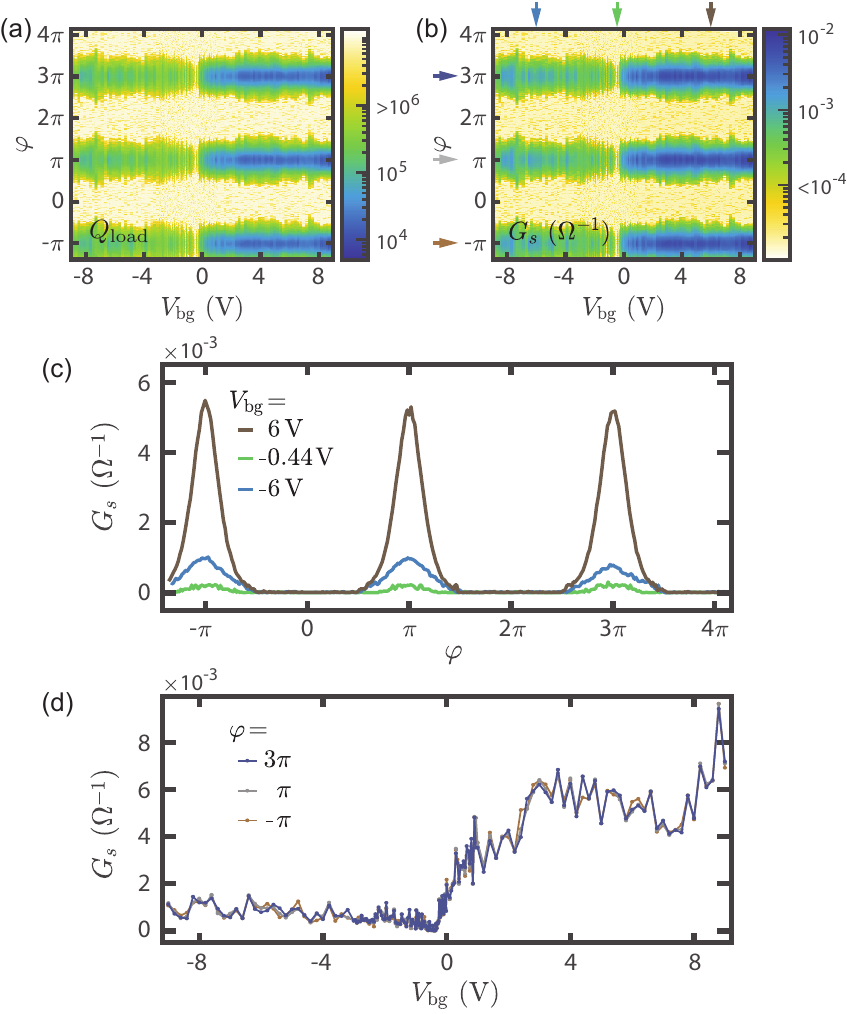}
\caption{\label{fig:loss} 
Evaluation of the shunt conductance $G_s$.
(a) The load quality factor $Q_{\rm{load}}$ in logarithmic scale as a function of $V_{\rm{bg}}$ and $\varphi$, deduced from resonance curve fittings.
(b) $G_s$ in logarithmic scale obtained by using Eq.\,\ref{eq:Qfull} with $Q_{\rm{load}}$ and the CPR results.
(c) Phase dependence of $G_s$ for different gate voltages.
(d) Gate dependence of $G_s$ for phase biasing conditions $\varphi=n_{\rm{odd}}\pi$.
}
\end{figure}

From Fig.\,\ref{fig:loss}(a), we observe that around the $0$-points ($\varphi=n_{\rm{even}}\pi$) the dissipation in the microwave circuit stemming from the rf SQUID is minor (${Q_{\rm{load}}>200\,000}$) for all $V_{\rm{bg}}$.
However, at the~$\pi$-points (${\varphi=n_{\rm{odd}}\pi}$), the dissipation becomes significantly larger and gate dependent with a minimal quality factor of ${Q_{\rm{load}}\approx 9800}$.

This behavior is reflected in $G_s$, which is mapped in Fig.\,\ref{fig:loss}(b) as a function of $V_{\rm{bg}}$ and $\varphi$.
Around the $0$-points, we deduce low conductance values \hbox{$G_{s}\leq0.1$~m$\Omega^{-1}$}, which refers to weak dissipation according to the parallel junction circuit model used here.
In contrast, at the $\pi$-points, a pronounced Lorentzian-shaped dissipation peak develops, as seen in Fig.\,\ref{fig:loss}(c).
The dissipation onsets are located symmetrically around the $\pi$-points and are weakly gate dependent.
On the other hand, the peak heights are strongly influenced by $V_{\rm{bg}}$ and reach a maximal value of $G_{s}\approx10$~m$\Omega^{-1}$ at large $n$-doping.
Although the amplitude of the peak appears to fluctuate as a function of $V_{\rm{bg}}$, the height replicates for the three different $\pi$-points measured here, as illustrated in Fig.\,\ref{fig:loss}(d).
This demonstrates the stability of the gate-tunable potential landscape in graphene.
In order to explain the dissipative response of the JJ, the underlying phase-dependent transport processes need to be consider, which are discussed in the next section.

\section{\label{sec:abs} Theory of Andreev Bound States}

In the following we relate the CPR and the phase-dependent dissipation to the microscopic concept of Andreev bound states (ABSs) formed within the JJ.

Coherent Andreev reflections of quasiparticles at the graphene-supercon\-ductor interfaces lead to the formation of ABSs~\cite{Andreev1964}.
These quasiparticle states transfer Cooper pairs across the junction in form of counter propagating electron-hole pairs~\cite{Pillet2010}. 
Due to the electron-hole symmetry, the ABSs come in pairs; one state has negative energy $E_n^{-} \leq 0$ and the other has positive energy $E_n^{+}= -E_n^{-}$, where $n$ denotes a specific transport channel.
The spectral gap $\delta E$ quantifies the minimal transition energy between states with negative and states with positive energies. 
Each occupied state carries current proportional to the derivative of its energy with respect to phase. 
The sum over the set of all channels defines the total supercurrent~\cite{beenakker1991universal}, which can be expressed as
\begin{equation}
\label{eq:cpr_diff}
I_s(\varphi)=\frac{2\pi}{\Phi_0}\sum_n f(E_n^{\pm}) \frac{\partial E_n^{\pm}}{\partial \varphi},
\end{equation}
where $f(E_n^{\pm})$ is a functional describing the occupation probability of the $n^{\rm{th}}$ ABS. 
In equilibrium the functional is given by the Fermi-Dirac distribution.

At zero temperature and in the absence of photons, all ABSs with negative energies are occupied \hbox{($f(E_n^{-})=1$)}, whereas all ABSs with positive energies are empty (${f(E_n^{+})=0}$).
In this situation the system is in the ground state and the occupation of the ABS spectrum is constant.
Therefore the supercurrent $I_s$ is free of any fluctuations. 
By virtue of the fluctuation-dissipation theorem,\,\cite{Hoffman1962} there is no dissipation and the effective junction shunt conductance assumes $G_s\rightarrow 0$.

When finite electronic temperatures $T$ and/or the absorption of photons from the electromagnetic environment are considered, the situation becomes different; 
thermal activation and/or microwave-induced transitions will drive the system out of the ground state.
The excitation-relaxation dynamics give rise to fluctuations in the ABS population, and correspondingly, in the supercurrent as well.
Consequently, there is dissipation and a finite shunt conductance $G_s$ appears~\cite{Martin-Rodero1996}.
When the spectral gap closes (${\delta E\rightarrow0}$) already small temperatures $T$ and small photon energies $hf$ will trigger fluctuations.
We note that the fluctuations are determined by the temperature, the photon absorption and emission rates and as well by the relaxation time $\tau_{\rm{rel}}$ of a non-thermal distribution towards a thermal one, which we express in the following as the energy  $\gamma=\hbar/(2\tau_{\rm{rel}})$.
In conclusion, this means that in general, both the inductive and dissipative part of a JJ depend on the ABS spectrum and the population dynamics within this spectrum.

Inherent to wide junctions -- like the graphene JJ investigated here -- is that there are various possible transport channels leading to many ABSs and hence to a dense ABS spectrum~\cite{Bretheau2017e}. 
The phase dependence of the ABS spectrum is determined by the geometry of the JJ and its material properties, i.e. the superconducting gap $\Delta$ in the leads and the inverse transport time in the normal region that relates to the Thouless energy $E_T$.
An important characteristic of JJs is whether they are in the `short' or `long' junction limit.
The former case is realized when $E_T \gg \Delta$, while the latter holds in the opposite limit.
The condition for the short junction limit can also be expressed as the coherence length $\xi$ being longer than the junction length $L$. 
Since this quantity for similar devices is reported to be $\xi\approx500$\,nm~\citep{Li2016,Bretheau2017e} and the junction under investigation has a length $L=400$\,nm, the condition for the short junction limit seems reasonably valid.

\begin{figure}[t!]
\includegraphics{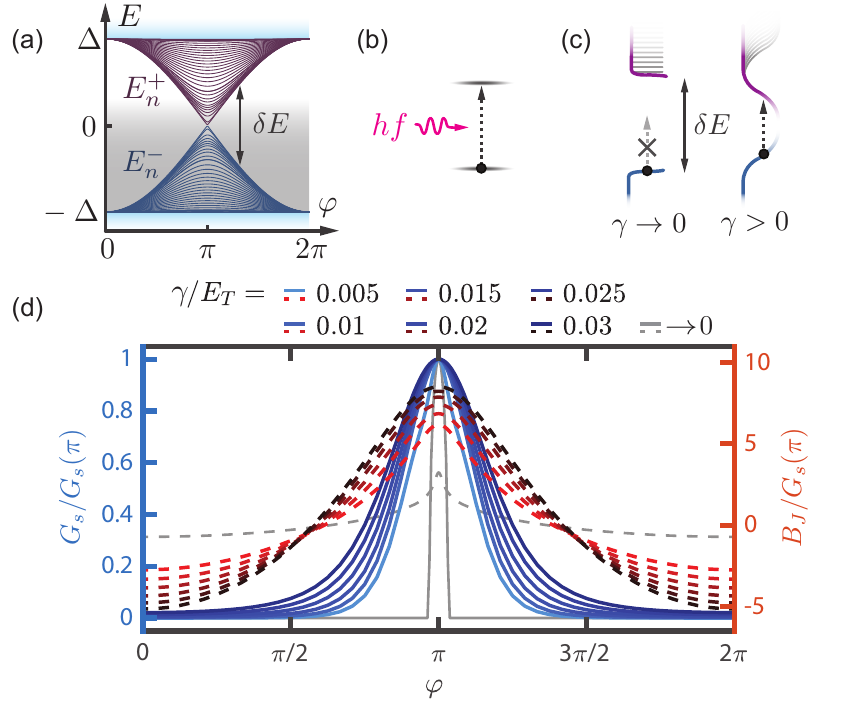}
\caption{\label{fig:spectrum} 
ABS spectrum and theoretical microwave response for a short, diffusive JJ.
(a) Spectrum of a short JJ with multiple channels of different transparencies.
(b) Microwave-induced transitions between states triggered by the absorption of a photon with energy $hf$.
(c) The finite lifetime of states described by the relaxation rate $\gamma$ causes a spectral broadening of the ABS energies and hence blurs the transition condition.
(d) Theoretically predicted dissipative and inductive response: $G_s$ (blue, left axis) and $B_J$ (red, right axis) normalized by the conductance value at $\varphi=\pi$ as function $\varphi$ for different $\gamma/E_T$ ratios. The normalization values for increasing $\gamma/E_T$ read: $G_s(\pi)/G_N=45,\,11,\,8.6,\,7,\,5.9,\,5,\,4.4,$\,where $G_N$ is the normal state conductance.
Here: $\Delta/E_T=0.1$, $hf/E_T=0.01$ and $kT/E_T=0.008$.
}
\end{figure}

For JJs in the short junction limit the ABS energies are given by 
${E_n^{\pm}(\varphi)=\pm\Delta\sqrt{1-\smash[b]{\tau_n\sin^2(\varphi/2)}}}$,
where $\tau_n$ is the trans\-mission probability of the $n^{\rm{th}}$ channel. 
Thus, the ABS spectrum strongly depends on the transparency distribution, which further defines the transport regime. 
For diffusive transport the transmission coefficients are continuously distributed following Dorokhov's bimodal distribution~\cite{Dorokhov1984},
which describes that there are many channels with low transmission ($\tau_n\rightarrow0$), but also many with high transmission probabilities ($\tau_n\rightarrow1$). 
Consequently, a dense ABS spectrum emerges as illustrated in Fig.\,\ref{fig:spectrum}(a) with a spectral gap ${\delta E=2\Delta|\cos(\varphi/2)|}$ that closes (${\delta E\rightarrow0}$) towards the $\pi$-points 
and maximally opens (${\delta E=2\Delta}$) towards the $0$-points.
In the following we assume predominant diffusive transport in the graphene JJ investigated here, which is supported by multiple observations: 
i) the small discrepancy between the experimentally determined skewness at large $n$-doping ($S\approx 0.22$) and the one predicted theoretically (${S=0.255}$)~\footnote{The CPR of a short, diffusive junction in equilibrium can be expressed analytically, from which one obtains a skewness ${S=0.255}$ at $T=0$~\cite{Kos2013,Heikkila} as indicated by the pink mark in Fig.~\ref{fig:cpr2}. The reduced skewness in the $p$-doped regime ($S\approx 0.12$) we assign to an overall suppression of the transmission probability due to the formation of $pn^\prime$-junctions at the graphene-superconductor interfaces.},
ii) the lack of Fabry-P\'{e}rot oscillations in the gate dependence of the CPR presented in Fig.~\ref{fig:cpr2} indicates suppressed ballistic transport~\cite{Nanda2017},
and iii) the randomly evolving shunt conductance $G_s$ seen in Fig.~\ref{fig:loss} hints at universal conductance fluctuations, which are expected for diffusive systems.
We believe that here the diffusive character of the device is stemming from scatting processes at the graphene edges, which are significant due to a small width to length ratio ($W/L\approx2$), and hence reduce the amount of ballistic channels.

In order to evaluate the dynamics of the ABS spectrum described above and translate it to lumped element quantities, we make use of theoretical works that predict the phase-dependent linear microwave response in terms of the susceptance $B_J$ and the shunt conductance $G_s$~\cite{Kos2013,Virtanen2011}.
For the theoretical analysis we consider a diffusive multi-channel JJ in the short junction limit at finite temperature coupled to a photonic environment of energy $hf$.
Note that in the experiment the photonic environment is provided by the driven microwave resonator.

First we consider the case of long-lived excitations ($\gamma\rightarrow0$), for which we find a sharp onset in $G_s(\varphi)$ as seen by the solid gray line in Fig.\,\ref{fig:spectrum}(d). 
The dissipation occurs in the phase range, where the spectral gap becomes smaller than the excitation energy ${\delta E \leq hf}$, thus allowing microwave-induced cross-gap transitions.
The width and the height of the dissipation peak depends on characteristic energy scales, which are denoted in the figure caption.
It is worth mentioning that not only transitions across the gap lead to dissipation; all possible absorption processes, including intra-band excitations $E_n^{+}\rightarrow E_m^{+}$, contribute to it, whereas the transition probability scales according to Fermi$^\prime$s Golden rule with the available density of states \cite{Dou2020}.
Fig.\,\ref{fig:spectrum}(b) depicts a microwave-induced transition of a quasiparticle from an arbitrary initial state to an available final state.
The fact that the ABSs have a finite lifetime causes a spectral broadening of the energies.
This results in a blurring of the transition condition (${\delta E\leq hf}$) as sketched in Fig.\,\ref{fig:spectrum}(c).
Therefore, increasing $\gamma$, i.e., shortening the lifetime, broadens the dissipation peak as seen by the blue lines in Fig.\,\ref{fig:spectrum}(d).
Importantly, the lifetime broadening also affects the susceptance, in particular the phase conditions for $B_J=0$ shift away from the $\pi$-point, which is equivalent to a reduction of the CPR skewness.
Note that $B_J$ for $\gamma\rightarrow 0$ shown in dashed gray appears different, because it is rescaled with a large conductance value $G_s(\pi)$.
A representation of Fig.\,\ref{fig:spectrum}(d) without normalization is shown in the SM~\cite{Note1}.

The influence of temperature on the microwave response is theoretically discussed, and together with experimental results, presented in the SM~\cite{Note1}. 

In short, environmental perturbations, namely, temperature and electromagnetic irradiation, cause dynamical variations in the population of ABS spectra on the timescale of the non-equilibrium occupation lifetime, which influence the susceptance $B_J$ likewise the CPR and give rise to dissipation captured by the shunt conductance $G_s$.


\section{\label{sec:compsimulation}Comparison with theory}

Finally, we compare the experimental results of the graphene junction with theoretical predictions based on the assumption of a short, diffusive multi-channel JJ. 

\begin{figure}[t!]
\includegraphics{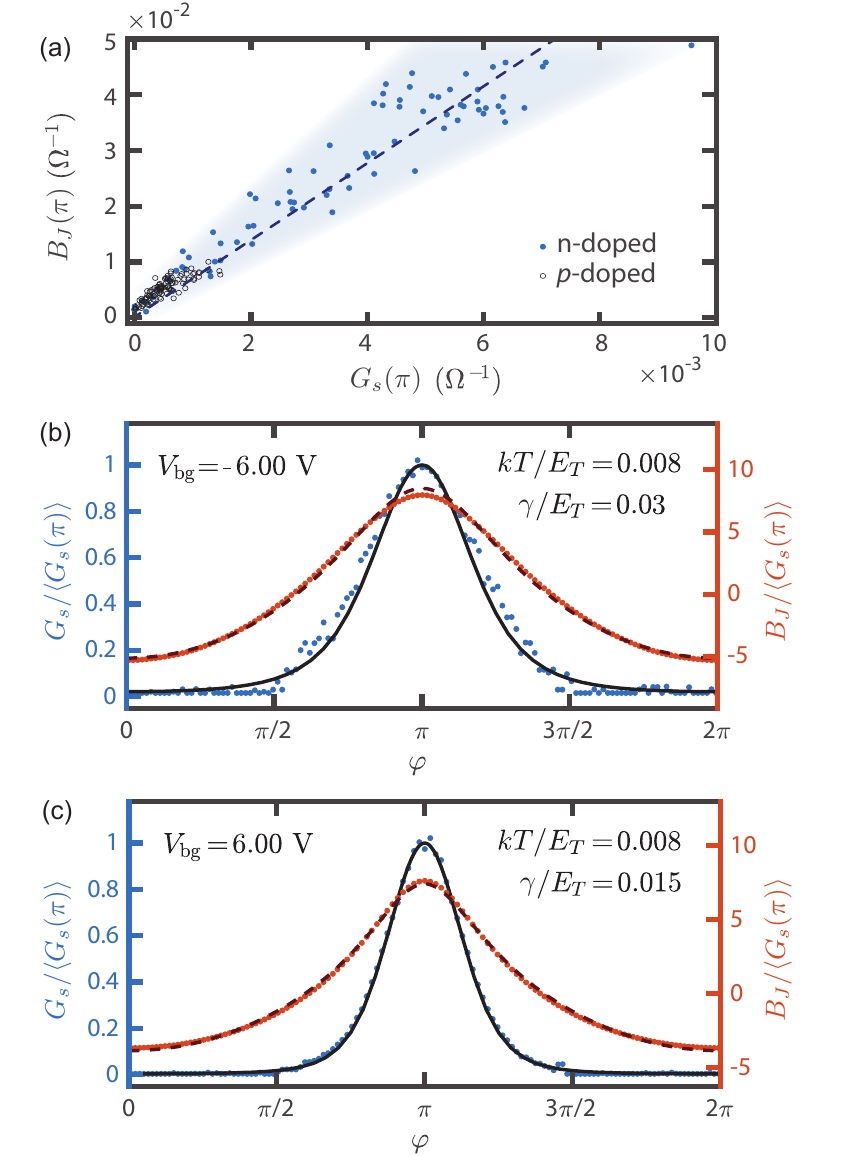}
\caption{\label{fig:comp}
Experimental observations in comparison with theoretical predictions for a short, diffusive JJ.
(a) Experimentally obtained susceptance $B_J$ versus shunt conductance $G_s$ at $\varphi=\pi$ follows a mean ratio of $\sim7$ indicated with the dashed line.
(b)-(c) Normalized measured $G_s$ (dotted blue, left axis) and $B_J$ (dotted red, right axis) overlaid with the normalized theoretical predictions for $G_s$ (solid) and $B_J$ (dashed), for which $\Delta/E_T=0.1$ and $hf/E_T=0.01$ are fixed, but $kT/E_T$ and $\gamma/E_T$ are variable. The best fitting parameter ratios are indicated. 
For $V_{\rm{bg}}=-6$\,V ($6$\,V) the normalizations read ${\langle{}G_s(\pi)\rangle=0.98\,{\rm{m}}\Omega^{-1}}$ (${5.23\,{\rm{m}}\Omega^{-1}}$) for the experimental traces and for the theoretical traces $G_s(\pi)/G_N=4.4\,(7)$. 
}
\end{figure}

One theoretical prediction, which was not explicitly pointed out above, is that the inductive and dissipative response ($B_J$,\,$G_s$) scale linearly with the normal state conductance $G_N$~\cite{Kos2013,Virtanen2011}, which is tunable with the gate voltage in our experiment.
From Fig.\,\ref{fig:comp}(a) one can verify this relation, since the relation between the experimentally deduced values of the susceptance $B_J$ and conductance $G_s$ obtained at $\varphi=\pi$ for all different $V_{\rm{bg}}$ -- clearly follows a linear trend.
Furthermore, the ratio $B_J/G_s$ is the inverse loss tangent describing the quality of the Josephson inductance~\cite{Lake2017}, where a larger ratio implies a more ideal behavior of the inductance.
We attribute the cone-shaped spread in Fig.\,\ref{fig:comp}(a) around the mean ratio ($\langle B_J(\pi)/G_s(\pi) \rangle\approx 7$) to altered ABS spectra and modified relaxation dynamics at different gate voltages. 

In the next step, we search for the best match between the theoretically predicted and the experimentally deduced phase-dependent microwave response by considering both the inductive and the dissipative properties of the JJ.
To this end, we numerically generate sets of $B_J$ and $G_s$ with different characteristic parameters. 
In particular, we vary the ratios $kT/E_T$ and $\gamma/E_T$ to account for a finite electronic temperature and to capture the effect of lifetime broadening.
We have fixed the Thouless energy to $E_T=10\Delta$ and the photon energy to $hf=\Delta/10$: the first condition ensures the short junction limit, whereas the second one compares favourably well to the expected experimental relation between the photon energy of the resonator and the superconducting gap of the contact material.

In Fig.\,\ref{fig:comp}(b) we compare the normalized theoretical and experimental values for $V_{\rm{bg}}=-6$\,V, whereas in Fig.\,\ref{fig:comp}(c) we perform the comparison for $V_{\rm{bg}}=6$\,V.
The experimental values $G_s$ (blue dots) and $B_J$ (red dots) are normalized with the shunt conductance at $\varphi=\pi$, denoted by $\langle G_s(\pi)\rangle$~\footnote{We average the three shunt conductance values closest to $\varphi=\pi$ to accommodate for scattering of the data.}.
Close overlap between theory and experiment can be found for both gate voltages with the same temperature ($kT/E_T=0.008$), but distinct relaxation rates $\gamma$.

At \hbox{$V_{\rm{bg}}=-6$\,V} we observe differences between the model and the experimental data even with the best match ($\gamma/E_T=0.03$). This is especially evident at the flanks of the dissipation peak and the susceptance at the $\pi$-point.
We attribute this mismatch to an inappropriate choice of transport regime for this gate voltage, because here the additional $pn^\prime$-junctions at the interfaces effectively elongate the quasiparticle trajectories. 
Consequently, the JJ tends to be in the long-junction limit causing a compression of the ABS spectrum.

On the other hand, we stress that we observe striking agreements between the theoretical predictions with $\gamma/E_T=0.015$ and the experimental data at $V_{\rm{bg}}=6$\,V.
Apparently, the model of a short, diffusive junction reproduces simultaneously the inductive and dissipative response of the graphene JJ for this doping configuration.
By evaluating the best fitting ratios $kT/hf=0.8$ and $\gamma/hf=1.5$ with the resonance frequency $f=3.098$\,GHz, we deduce an electronic temperature $T=120$\,mK and obtain a relaxation time $\tau_{\rm{rel}}=17$\,ps.
A similar equilibration time ($\tau_{\rm{rel}}=7$\,ps) is reported for an equivalent short, diffusive Al-graphene JJ probed at mK temperatures and large $n$-dopings~\cite{Voutilainen2011}.
We stress that the ABS spectrum of a short, diffusive junction might not be the only spectrum, which in a similar theoretical model could reproduce the experimentally observed response.
In particular, in a wide JJ the ABS spectrum can be built from quasiparticles with long and short trajectories leading to more complex ABS structures than discussed above~\cite{Bretheau2017e}.


\section{\label{sec:conclusion}Conclusion}

We have measured the reflective response of a microwave resonator inductively coupled to a graphene-based rf SQUID as a function of flux-bias and charge carrier density. 
We developed a concise circuit model to infer the CPR and the phase-dependent dissipation of the graphene JJ from the changes in the resonance frequency and broadening.
We hereby obtain the full complex admittance of the junction, which is the key parameter to design Josephson microwave circuits.

Our comprehensive investigation demonstrates the impact of the environment on the performance of JJs in terms of finite temperature and microwave photons. 
If the environment provides energies larger than the spectral gap, short-lived excitations appear in the ABS spectrum, which induce fluctuations in the supercurrent leading to dissipation.
The comparison between the experimentally deduced microwave response at high electron density and the one predicted by theory for a short and diffusive junction model, yields striking agreement, from which we deduce a relaxation time of 17\,ps.
This fast thermal relaxation makes graphene-based JJs unique candidates for highly sensitive and fast bolo- and calorimeters\,\cite{Efetov2018,Lee2020,Kokkoniemi2020}.

Furthermore, the device architecture and measurement protocols presented in this work are well-suited to explore the fundamental properties of other JJs, such as junctions made of 2D/3D topological insulators or Dirac and Weyl semimetals~\cite{Murani2019}.
Particularly, the topological nature of these JJs can be probed, because it is predicted that they host ABS states that cross at the $\pi$-points but possess opposite parities, meaning that microwave-induced transitions across the gap are prohibited \cite{Peng2016a}. 
As a consequence, it is expected that the dissipative character of topological JJs is distinctly different from trivial ones~\cite{Lutchyn2010,Dmytruk2016a,Trif2018b}.


\begin{acknowledgments}

We thank S.~Dehm for technical support in the nanofabrication facility at KIT.
We are grateful for discussions about general properties of 2D Josephson junction with A.~Kononov and P.~Karnatak.
This research was supported by the Swiss National Science Foundation through a) grants No 172638 and 192027, b) the National Centre of Competence in Research Quantum Science and Technology (QSIT), and c) the QuantEra project SuperTop; the János Bolyai Research Scholarship of the Hungarian
Academy of Sciences, the National Research Development and Innovation Office (NKFIH) through the OTKA Grants FK 132146 and NN127903 (FlagERA Topograph), and the National Research, Development and Innovation Fund of Hungary within the Quantum Technology National Excellence Program (Project
Nr. 2017-1.2.1-NKP-2017-00001), the Quantum Information National Laboratory of Hungary and the ÚNKP-20-5 New National Excellence Program. We further acknowledge funding from the European Union’s Horizon 2020 research and innovation programme, specifically a) from the European Research Council (ERC) grant agreement No 787414, ERC-Adv TopSupra, and b) grant agreement No 828948, FET-open project AndQC.
This work was partly supported by Helmholtz society through program STN and the DFG via the projects DA 1280/3-1.
K.~Watanabe and T.~Taniguchi acknowledge support from the Elemental Strategy Initiative
conducted by the MEXT, Japan, Grant Number JPMXP0112101001, JSPS
KAKENHI Grant Number JP20H00354 and the CREST(JPMJCR15F3), JST.

All raw- and metadata in this publication are available in numerical form  together with the processing codes at DOI:~\href{https://doi.org/10.5281/zenodo.4479896}{10.5281/zenodo.4479896}.
\end{acknowledgments}

%

%
\pagebreak
\clearpage
\widetext

\setcounter{equation}{0}
\setcounter{figure}{0}
\setcounter{table}{0}
\setcounter{page}{1}
\setcounter{section}{0}

\renewcommand{\thefigure}{S\arabic{figure}}
\renewcommand{\theequation}{S\arabic{equation}}
\renewcommand{\thesection}{S\Roman{section}}
\renewcommand{\bibnumfmt}[1]{[S#1]}
\renewcommand{\citenumfont}[1]{S#1}

\textbf{\centering\large Supplementary Material:\\}
\textbf{\centering\large Phase-dependent microwave response of a graphene Josephson junction\\}

\vspace{1em}


\section{\label{sec:fabrication}Fabrication}
\subsection{NbTiN sputtering}
The NbTiN film (80\,nm)  is sputtered on an undoped Si/SiO$_2$ wafer ($500\,\mu$m/170\,nm) in a AJA$^{\copyright}$ ATC Orion 8 sputtering machine using a NbTi-target (70/30 at\%, 99.995\% purity) and N$_2$ added to the Ar sputtering gas. 
Before deposition, following wafer cleaning steps are performed:
\begin{itemize}
\setlength\itemsep{0em}
\item{10 min sonication in deconex$^{\copyright}$ 12 basic/DI-water solution $\rightarrow$ flush with DI-water}
\item{10 min sonication in DI-water $\rightarrow$ blow-dry}
\item{~5 min baking at 120\,$^\circ$C}
\item{10 min sonication in acetone}
\item{10 min sonication in IPA $\rightarrow$ blow-dry}
\item{~5 min UV-ozone in UVO-Cleaner$^{\copyright}$ Model 42-220}
\end{itemize}
After inserting the wafer into the sputtering machine typical base pressures of $\sim8\times10^{-9}$\,Torr are achieved.
We position the wafer as close as possible to the NbTi target to minimize particle scattering.
Before depositing on the wafer substrate -- the chamber and the NbTi-target are conditioned. During the conditioning steps a substrate shutter protects the wafer from material deposition. 
We pre-sputter Ti (35\,sccm of Ar at 4\,mTorr with (DC) 100\,W for 20\,min) to remove oxygen residuals in the chamber. 
After terminating the Ti pre-sputtering, we pump on the chamber until pressures $<1\times10^{-9}$\,Torr are reached, which typically takes $\sim20$\,min. Then we sputter  NbTi+N$_2$ (50\,sccm of Ar, 3.5\,sccm of N$_2$ at 2\,mTorr with (DC) 275\,W). 
After 4\,min of sputtering time, we open the substrate shutter to allow for deposition on the wafer -- for totally 375\,s, which results in a film thickness of $\sim80$\,nm. 
The N$_2$-flow was optimized separately to achieve a stoichiometric compound of NbTiN. 


\subsection{Resonator and graphene Josephson junction}
After the sputter deposition of NbTiN -- the resonator is defined, then the separately prepared van der Waals heterostructure is placed and contacted as explained in the following:

\textbf{Resonator patterning:} The resonant structure and supply lines in NbTiN are patterned by using positive e-beam lithography (EBL) followed by an Ar/Cl$_2$ reactive-ion etching-step. 
The meandered co-planar transmission line (TL) is designed with a central conductor width of 12~$\mu$m, a clearance of 6~$\mu$m to the surrounding ground plane and a total length of $l=7.54$\,mm.  
The TL is shorted to ground on one side and interrupted by a coupling capacitance on the other, which forms the $\lambda/4$-resonator.
The dimensions of the finger capacitor are aimed for providing a coupling capacitance of $\sim4$\,fF.
From the measurement and calculations presented below we deduce for this configuration a resonance frequency $f_0\approx3.098$\,GHz, a characteristic impedance $Z_r=69.5\,\Omega$ and a coupling capacitance $C_c=4.7$\,fF.
The supply lines for the gate, flux, and pump are designed on purpose with different aspect ratios to provoke an impedance mismatch for minimizing loss channels for the resonator.
For this experiment we do not make use of the pump line (upper right in Fig.\,\ref*{fig:device}(a) of the maintext).
The fabrication details are listed below. 

\begin{itemize}
\setlength\itemsep{0em}
	\item Resist: AR-P$^{\copyright}$ 671.05 (positive PMMA resist, 950k, 5$\%$ diluted in chlorobenzene) 
	\item Coating: Spin with 6000\,rpm for 40\,s (ramp 4\,s) $\rightarrow$ thickness $600$\,nm. 
	\item Bake: 5\,min on the hot plate at 180\,$^\circ$C 
	\item EBL with Zeiss$^{\copyright}$ Supra 40: 20\,kV acceleration voltage, aperture of 60\,$\mu$m in high current mode and a dose of 275\,$\mu$C/cm$^2$
	\item Resist development: MIBK/IPA 1:3 at room temperature for 60\,s. The development is followed by 10\,s in IPA $\rightarrow$ blow-dry
	\item O$_2$-plasma in Oxford$^{\copyright}$ Plasmalab 80 Plus: 2\,min with 30\,W and 16\,sccm of O$_2$-flow at 250\,mTorr
	\item NbTiN etching in Sentech$^{\copyright}$ SI500: 1\,Pa, ICP power 100\,W, RF power 125\,W with 40\,sccm of Cl$_2$ and 25\,sccm of Ar for 40\,s; etch rate $\sim3$\,nm/s
	\item Lift-off: 1\,h in acetone at 50\,$^\circ$C followed by sonication and rising with IPA $\rightarrow$ blow-dry 
\end{itemize}

\textbf{Preparation of the graphene van der Waals heterostructure:} The graphene Josephson junction (JJ) is made from a van der Waals heterostructure, which consists from bottom to top out of a thick graphite sheet, a bottom hexagonal boron nitride (hBN) with thickness $d=47.5$\,nm,  a monolayer graphene and a top hBN (21\,nm).
We separately prepare the stack by standard polycarbonate-assisted pick-up technique~\cite{S_Zomer2014} and place it next to the current anti-node of the resonator. 
Details about the stacking routine is provided in the SI of Ref.~\cite{S_Indolese2020}. 
In the end of this process the whole device is placed for 1\,h in dichlormethane to dissolve polycarbonate residuals to prepare the stacks surface for the following fabrications steps.

\textbf{Contacting and Shaping:} The graphene is contacted and enclosed by a thermally evaporated Ti/Al (5/90\,nm) lead, which forms the rf SQUID. Access regions for the self-aligned side contacts~\cite{S_Wang2013} are structured with positive e-beam lithography in combination with CHF$_3$/O$_2$ etching. 
The contacts to the graphene and the loop are fabricated simultaneously, which was done at the Institute of Nanotechology of the Karlruhe Institute of Technology (KIT).
Rainer Kraft from KIT guided the fabrication, for which we made use of following recipe~\cite{S_Kraft2018}:
\begin{itemize}
\setlength\itemsep{0em}
	\item Resist: AR-P$^{\copyright}$ 672.045 (positive PMMA resist, 950k, 4.5$\%$ diluted in anisole)
	\item Coating:  Spin with 300\,rpm for 2\,s (ramp 1.5\,s) followed by 6000\,rpm for 60\,s (ramp 1.5\,s) $\rightarrow$  thickness 280\,nm
	\item EBL: 30\,kV acceleration voltage and a dose of 360\,$\mu$C/cm$^2$. The 20\,$\mu$m (120\,$\mu$m) aperture was used for small (large) structures.
	\item Resist development: MIBK/IPA 1:3 at room temperature for 20s. The development is followed by 2\,s in IPA $\rightarrow$ blow-dry
	\item Opening hBN windows for edge-contact: Plasma etching in Oxford$^{\copyright}$ Plasmalab 80 Plus: CHF$_3$/O$_2$ 40\,sscm/4\,sccm with 60\,W at 60\,mTorr.
	The etch time is adjusted to the thickness of the top hBN with calibrated etch rate of 0.55\,nm/s.
	\item Thermal evaporation of Ti/Al contacts in Bestec using the same mask:
	    \begin{itemize}
       		\item \underline{Ti:}  $T_{\rm{source}}=1640\,^\circ$C; ramp rate 25\,$^\circ$C/min; pressure $7.8\times 10^{-9}$\,mbar;\\
       		 evaporation rate 0.5\,nm/min; film thickness 5\,nm, $T_{\rm{stage}}=-130^\circ$C
       		\item \underline{Al:} $T_{\rm{source}}=1200\,^\circ$C; ramp rate 25\,$^\circ$C/min; pressure $3.5\times 10^{-9}$\,mbar;\\ 
			evaporation rate 6\,nm/min; film thickness 90\,nm, $T_{\rm{stage}}-130^\circ$C
			\item The Bestec at KIT has a specially large ratio between source-diameter and source-sample-distance, which is providing substantial undercut deposition.
     	\end{itemize}
	\item Lift-off: 2\,h in acetone at room temperature, afterwards rinsed with IPA and blow dried with N$_2$
\end{itemize}

After contacting, the graphene stack is shaped to a width $W=1\,\mu$m using a positive PMMA resist mask in combination with CHF$_3$/O$_2$ etching step. This process step is equivalent to the one described above.

\textbf{Design of the rf SQUID:} For the loop material it is desirable to chose a material with low kinetic inductance, because this adds beside the geometric inductance to the total self-inductance of the loop $L_{\rm{loop}}$. 
If $L_{\rm{loop}}I_c>\Phi_0/(2\pi)$, where $\Phi_0$ is the flux quantum and $I_c$ is the critical current of a sinusoidal current-phase relation, then the phase $\varphi$ across the junction becomes hysteretic as a function of external flux due to screening effects, such that the phase condition ${\varphi=\pi}$ cannot be reached.
Therefore we chose Al, which has a low kinetic inductance and a relatively small superconducting gap $\Delta_{\rm{Al}}=180\,\mu$eV advantageous for keeping the critical current low. 
The elongated shape of the loop and the varying lead width (see Fig.\,\ref*{fig:device} (b) in maintext) builds a compromise between maximizing the coupling to the resonator and minimizing screening effects. For this specific geometry we find a self-inductance $L_{\rm{loop}}=211$\,pH from finite-element simulations performed in Sonnet~\cite{S_manual2005sonnet}, in which we assume a kinetic sheet inductance of $0.2$\,pH/$\square$ for the 90\,nm Al film calculated from the value presented in Ref.~\cite{S_Hu2020}. Simultaneously, we obtain the mutual inductance $M=30.83$ pH between the resonator and the SQUID loop. For both of these results the exact geometry of the device deduced by scanning electron microscope imaging was considered. 

The loop surrounds an area $A\approx4\,000$\,$\mu$m$^2$ implying that already $\sim1~\mu$T generates a flux quantum inside or in other words provides one full phase biasing period. This small magnetic field is governed by the DC current $I_{\rm{flux}}$ flowing close by the loop. 


\newpage
\subsection{Bonding}
\begin{figure}[!h]
\includegraphics[width=1\columnwidth]{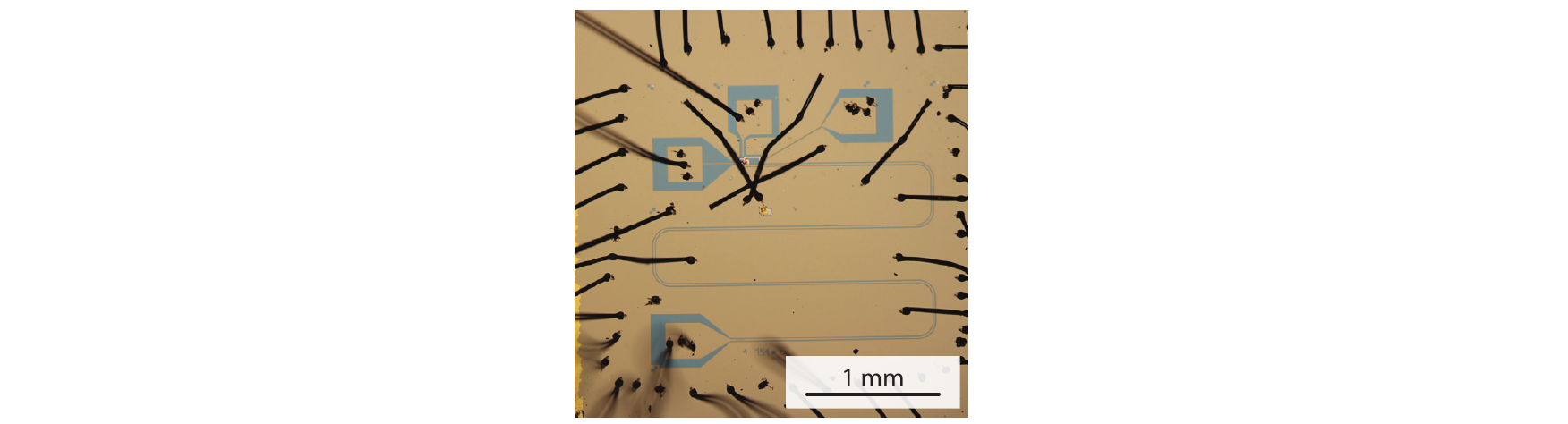}
\caption{\label{fig:bonding}
Optical picture for illustrating the bonding of the device.
}
\end{figure}

After gluing the sample with silver paste onto the copper backplane of the PCB holder, we connect the rf and DC lines from the PCB with Al bond wires to the once of the device. We ensure a homogeneous ground plane by adding many grounding bonds around the resonant structure. Additionally, we place bond-bridges across the TL and between the areas surrounding the rf SQUID. Fig.\,\ref{fig:bonding} presents the bond arrangement used for this devices. Note that this pictures is taken after removing the sample from the PCB holder after the measurements and some bonds broke off.


\section{\label{sec:setup}Set-up overview}

\begin{figure}[!h]
\includegraphics[width=1\columnwidth]{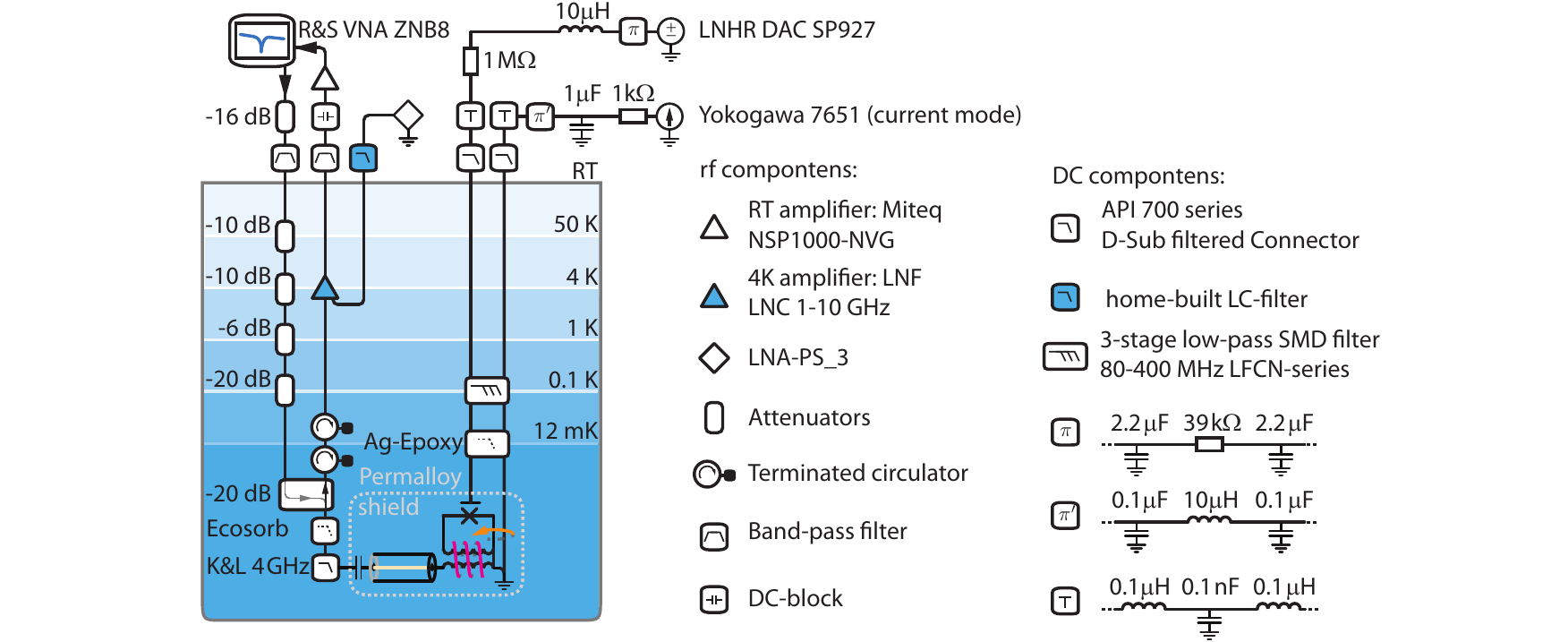}
\caption{\label{fig:setup}
Detailed overview of the measurement set-up.
}
\end{figure}

The measurements are carried out in a BlueFors$^\copyright$ BF-LD400 cryogen-free dilution refrigerator, in which the mixing chamber plate is modified with an additional mounting stage for high frequency components. 
A detailed overview of the high frequency and DC set-up is provided in Fig.\,\ref{fig:setup}.
The device is surrounded by a permalloy shield to screen external magnetic field fluctuations. 
We probe the resonant structure with a vector network analyser (VNA) in a standard reflectometry configuration. 
The probe signal reaches the sample via an attenuated input line and a directional coupler. The reflected signal travels back to the VNA through the amplification chain consisting of a room temperature amplifier and 4\,K-HEMT amplifier, which are isolated towards the device by two circulators located at the mixing chamber plate. 
The supply lines for the gate and bias voltages of the 4\,K-HEMT amplifier are filtered with home-built $LC$-lowpass filters (cut-off $\sim150$\,kHz).
The DC lines for tuning the gate voltage on the graphene and the flux inside the rf SQUID are heavily filtered at room and base temperatures.
The Ecosorb$^\copyright$ lowpass-filter in the high frequency line does have a cut-off frequency around 13\,GHz and the silver-epoxy lowpass-filters in the DC lines do have a cut-off around 6\,MHz.


\newpage
\section{\label{sec:readout}Read-out}
\begin{figure}[!h]
\includegraphics[width=1\columnwidth]{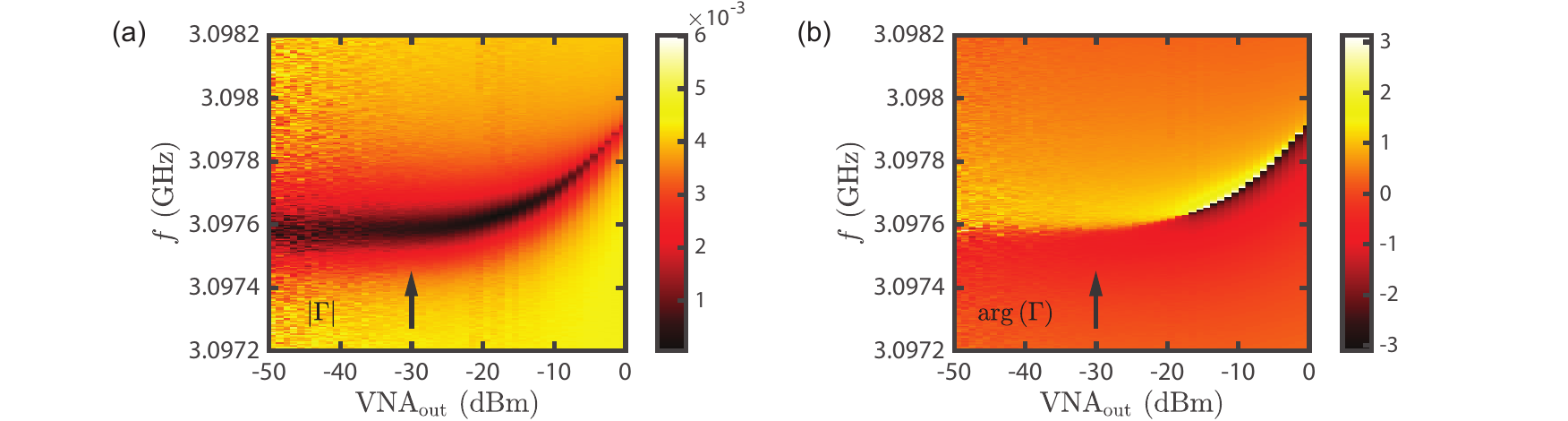}
\caption{\label{fig:powerdep} Reflection coefficient $\Gamma$ at $V_{\rm{bg}}=4.5$\,V and $\varphi=\pi$ as a function of probe frequency $f$ and probe power VNA$_{\rm{out}}$ obtained with a bandwidth VNA$_{\rm{BW}}=$50\,Hz. (a) Amplitude and in (b) argument of $\Gamma$. All subsequent measurements are carried out with VNA$_{\rm{out}}=-30$\,dBm (indicated with the arrows.)
}
\end{figure}

The read-out power has a substantial influence on the response of the coupled resonator rf SQUID circuit. 
In Fig.\,\ref{fig:powerdep} we present the reflectance coefficient $\Gamma$ as a function of probe frequency $f$ and probe power VNA$_{\rm{out}}$. 
Here, the graphene JJ is gated with $V_{\rm{bg}}=4.5$\,V (electron doped) and tuned to $\varphi=\pi$ (spectral gap is smallest). 
We observe that for increasing the probe power the resonance frequency shifts to higher values and the resonance lineshape alters.
We attribute this to non-linear effects caused by over driving the resonator or saturating the ABS spectrum. 
It could also be that the stray field of the resonator induces large phase biasing oscillations, which smears out the phase-dependent features of the JJ. 
Additionally, irradiating JJs affect their current-phase relation~\cite{S_Basset2019} and the $IV$-characteristics develops Shapiro steps~\cite{S_Shapiro1963}. Both of the effects will influence the reflective response.

If the read-out power is below -25\,dBm, there are no more changes in the resonance frequency nor in the resonance lineshape. All subsequent measurements are carried out with VNA$_{\rm{out}}=-30$\,dBm and a bandwidth VNA$_{\rm{BW}}=$500\,Hz.

\subsection{\label{subsec:photonn}Conversion to photon number}
The averaged photon number in the resonator can be estimated with following expression~\cite{S_Burnett2018}
\begin{equation}
\label{eq:photonnumber}
\langle n\rangle=\frac{2}{\hbar\omega_0^2}\frac{Z_0}{Z_r}\frac{Q^2}{Q_c}P_{\rm{app}},
\end{equation}
where $\hbar$ is the reduced Planck constant, $Z_0$ is the environmental impedance, $Z_r$ is the characteristic impedance of the co-planar transmission line, $Q$ is the total quality factor of the resonant structure, $Q_c$ is the coupling quality factor and $P_{\rm{app}}$ is the applied microwave power.
With $\omega_0=2\pi\cdot3.1$\,GHz, $Z_0=50\,\Omega$, $Z_r=69.5\,\Omega$, $Q=24\,000$, $Q_c=24\,000$ and $P_{\rm{app}}=-130$\,dBm($=10^{-16}$\,W) we obtain an intra cavity photon occupation $\langle n\rangle\approx90$. However this value can deviate from the actual photon number, since $P_{\rm{app}}$ is only estimated via the total attenuation measured at room temperature in combination with the output power of the VNA.


\subsection{\label{subsec:probeflux}Probe flux $\delta \Phi$}
\begin{figure}[!h]
\includegraphics[width=1\columnwidth]{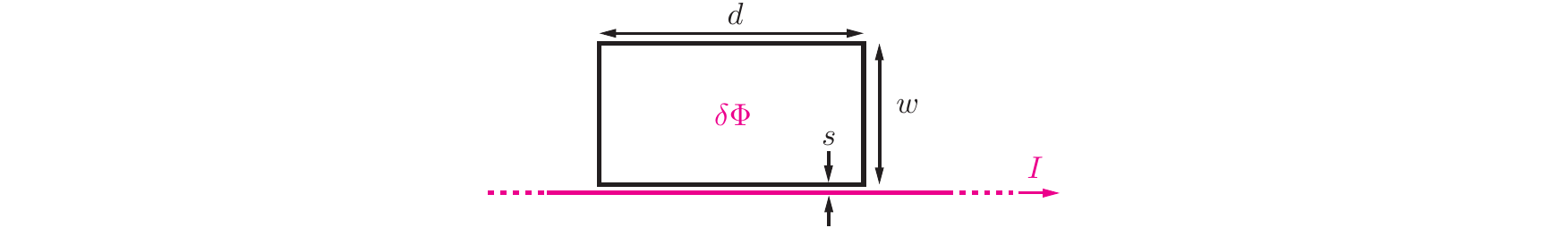}
\caption{\label{fig:biotsavart} Rectangular loop next to straight current-carrying wire.
}
\end{figure}

In order to estimate the probe flux $\delta \Phi$ we first evaluate the current at the end of the transmission line (TL) and then translate the current to the magnetic field strength. 
From TL theory~\cite{S_pozar2011microwave} we can derive a formula, which is expressing the current at the shorted end of a lossless TL capacitively coupled to a generator
\begin{equation}
I_{\rm{TL}}(f)=\frac{V_{\rm{gen}}}{\sin(\beta l)}\frac{\tan(\beta l)}{jZ_r\tan(\beta l)+Z_0+(j2\pi fC_c)^{-1}},
\end{equation}
where $V_{\rm{gen}}=10^{\frac{P_{\rm{app}}[{\rm{dBm}}]-10}{10}}[{\rm{V}}]$ is the generator voltage, $Z_r$ is the characteristic impedance of the TL, $Z_0$ is the input impedance of the generator, $\beta=2\pi f\frac{\sqrt{\epsilon_{\rm{eff}}}}{c}$ is the wavenumber of the TL, $\sqrt{\epsilon_{\rm{eff}}}$ is the effective refractive index, $l$ is the length of the TL and $C_c$ is the coupling capacitance.  By maximizing the absolute value of this expression for the frequency, one obtains the maximal current in the TL provided at resonance.
With $P_{\rm{app}}=-130$\,dBm, $\epsilon_{\rm{eff}}=10.24$, $Z_0=50\,\Omega$, $Z_r=69.5\,\Omega$, $l=7.54$\,mm and $C_c=4.7$\,fF we obtain $|I_{\rm{TL}}|_{\rm{max}}=310$\,nA at 3.094\,GHz.
The Biot-Savart law expresses the magnetic field magnitude $B$ at distance $r$ apart from a long, thin wire, carrying a steady current in free space
\begin{equation}
B=\frac{\mu_0}{2\pi}\frac{I}{r},
\end{equation}  
where $\mu_0=4\pi\cdot10^{-7}~\mathrm{N/A^{-2}}$ is the vacuum permeability. 
By substituting values $I=310$\,nA and $r=1~\mathrm{\mu{}m}$ one gets $B=62$~nT. 
Furthermore, we are interested in the flux created by this current within a rectangular loop, which can be expressed by
\begin{equation}
\delta\Phi=\frac{\mu_{0}I}{2\pi}\cdot{}d\cdot{}\ln{\left(\frac{s+w}{s}\right)},
\end{equation}  
where $d$ is the length of the loop, $w$ is the width of the loop and $s$ describes the spacing from the wire to the closer loop edge see Fig.\,\ref{fig:biotsavart}. 
With $I=310\,\mathrm{nA}$, $d=80\,\mathrm{\mu{}m}$, $w=40\,\mathrm{\mu{}m}$ and $s=1\,\mathrm{\mu{}m}$, which roughly mimics the dimensions of the rf SQUID, we obtain $\delta\Phi\approx0.01\,\Phi_0$, where $\Phi_0\approx2\times10^{-15}$\,Wb is the magnetic flux quantum.

\section{\label{sec:rescurvefit}Resonance curve fitting}
\begin{figure}[!h]
\includegraphics[width=0.75\columnwidth]{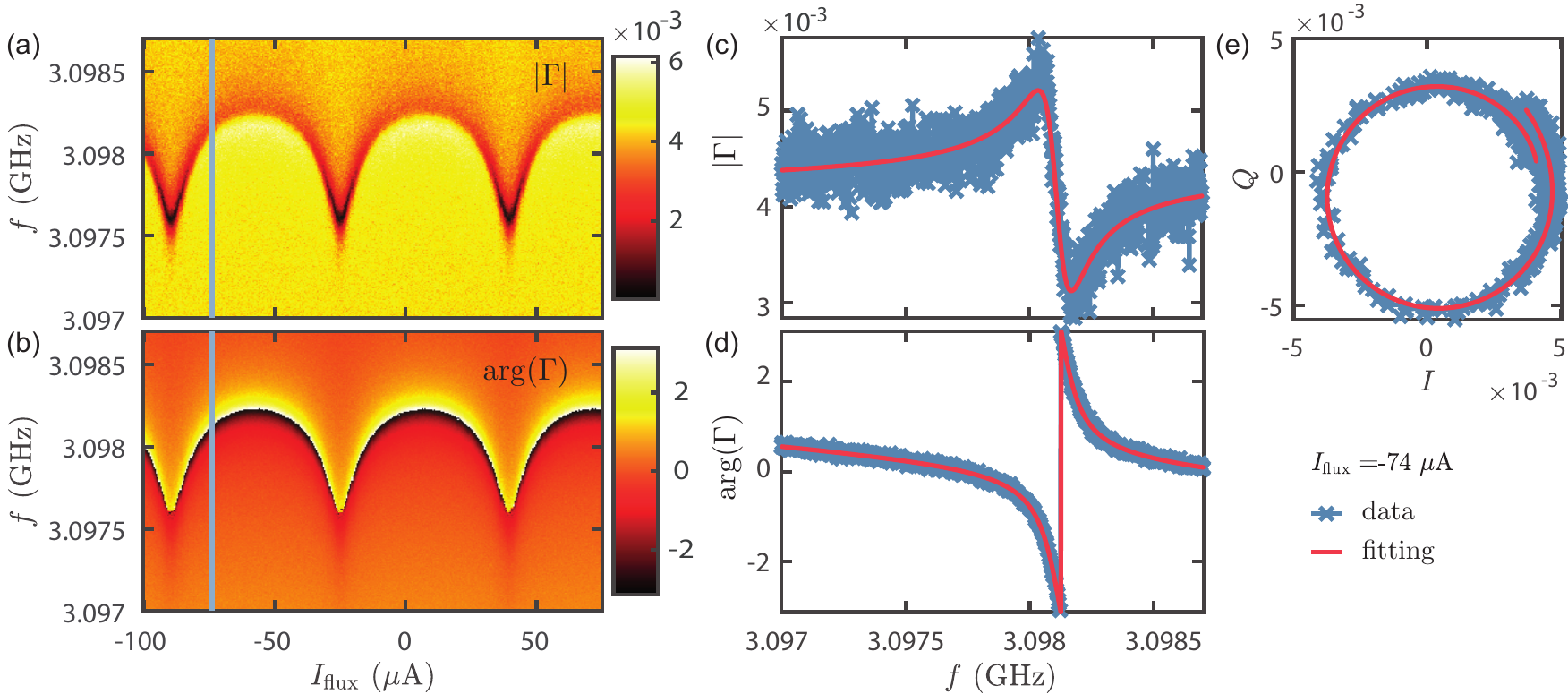}
\caption{\label{fig:rescurvefitting}
Flux dependence of the reflection coefficient $\Gamma$ at $V_{\rm{bg}}=5$\,V. (a)-(b) Colormaps of $|\Gamma|$ and $\arg(\Gamma)$ as a function of probe frequency $f$ and DC flux current $I_{\rm{flux}}$. (c)-(d) Resonance curve in $|\Gamma|$ and $\arg(\Gamma)$ at $I_{\rm{flux}}=-74\,\mu$A. (e) Resonance curve in the $IQ$-plane. (c)-(e) The measured data is shown as blue crosses, while the fit result is presented as solid red line. Following parameters are deduced from fitting:}
\begin{ruledtabular}
\begin{tabular}{cccccccc}
\textrm{$f_0$}&
\textrm{$\phi$}&
\textrm{$Q_c$}&
\textrm{$Q_e$}&
\textrm{$a_{\rm{off}}$}&
\textrm{$a_{\rm{slope}}$}&
\textrm{$p_{\rm{off}}$}&
\textrm{$p_{\rm{slope}}$}\\
\colrule
\vspace*{-0.2cm}
\\
$3.0981$\,GHz & $0.25$\,rad & $23\,800$ & $669\,800$ & $4.3\times10^{-3}$ & $5.1\times10^{-11}$/Hz & $0.16$\,rad & $-4.7\times10^{-7}$\,rad/Hz\\
\end{tabular}
\end{ruledtabular}
\label{fig:reflectance_curve}
\end{figure}

Changing the flux biasing ($I_{\rm{flux}}=\left[-100,75\right]~\mu$A) in the graphene rf SQUID coupled to the $\lambda/4$-resonator influences the resonant behavior of the circuit as seen in the reflectance curve maps ($V_{\rm{bg}}=5$\,V) presented in Fig.\,\ref{fig:rescurvefitting}(a) and Fig.\,\ref{fig:rescurvefitting}(b).
We implement a fitting routing, which is taking into account both the amplitude and the argument of $\Gamma$ at once to insure a highly robust fitting procedure.
An other advantage of the method is the clear distinction between the coupling quality factor $Q_{\rm{c}}$ and the effective quality factor $Q_{\rm{e}}$.
In the following we consider the resonance curve obtained at $I_{\rm{flux}}=-74~\mu$A. From Fig.\,\ref{fig:rescurvefitting}(c) we observe that $|\Gamma|$ has a shallow asymmetric lineshape and from Fig.\,\ref{fig:rescurvefitting}(d) we observe that $\arg(\Gamma)$ develops a $2\pi$-jump.
In the $IQ$-plane, where $I=\operatorname{Re}(|\Gamma|e^{j\arg{(\Gamma)}})$ and $Q=\operatorname{Im}(|\Gamma|e^{j\arg{(\Gamma)}})$, the resonance curve generates here a circle surrounding the $IQ$-point=(0,0) as
shown in Fig.\,\ref{fig:rescurvefitting}(e).
We fit both $|\Gamma|$ and $\arg{(\Gamma)}$ simultaneously with a least-square method with following combination of formulas
\begin{equation}
\label{eq:complex_fit_SM}
\Gamma= \left[\frac{\Gamma_{\rm{min}}+2j{}Q\frac{f-f_0}{f_0}}{1+2j{}Q\frac{f-f_0}{f_0}}-1\right]e^{j\phi}+1,
\end{equation}
where $\Gamma_{\rm{min}}=\frac{Q_c-Q_e}{Q_c+Q_i}$ is the minimal reflection coefficient in the symmetric case ($\phi=0$), $Q=(Q_c^{-1}+Q_e^{-1})^{-1}$ is the total quality factor, $Q_e=(Q_i^{-1}+Q_{\rm{load}}^{-1})^{-1}$ is the effective quality factor, in which $Q_{\rm{load}}$ is the quality factor of the load, $f$ is the probe frequency, $f_0$ is the resonance frequency and $\phi$ is the asymmetry angle, which causes a rotation of the resonance circle in the $IQ$-plane around the off-resonance point. 
In order to account for an offset and a slope in $|\Gamma|$ as well as in $\arg{(\Gamma)}$, we make use of following expression, which together with Eq.\,\ref{eq:complex_fit_SM} provides the complete fitting formula:  
\begin{equation}
\label{eq:complex_fit:code}
\Gamma_{\rm{fit}}=|\Gamma|\cdot[a_{\rm{off}}+a_{\rm{slope}}(f-f_0)]\cdot e^{j[\arg{(\Gamma)}+p_{\rm{off}}+p_{\rm{slope}}(f-f_0)]},
\end{equation}
where $a_{\rm{off}}$ describes an offset in the amplitude, $a_{\rm{slope}}$ describes a slope in the amplitude, $p_{\rm{off}}$ describes an offset in the argument and  $p_{\rm{slope}}$ describes a slope in the argument.
In Fig.~\ref{fig:rescurvefitting} the fit result (solid red) is overlain with the measurement data (blue crosses) and the complete set of fitting parameters is listed.
\newpage


\section{\label{sec:loadedres}Loaded resonator}
The performance of a resonator is depending on the load impedance $Z_{\rm{load}}$ attached to it. In order to relate the resonance frequency $f_0$ and the quality factor $Q$ of the resonator to properties of $Z_{\rm{load}}$, one can compare the input impedance of the specific circuit with the one of a known circuit.
Here, we will compare a loaded $\lambda/4$-resonator with a parallel $RLC$-circuit.
 \begin{figure}[!b]
\includegraphics[width=0.8\columnwidth]{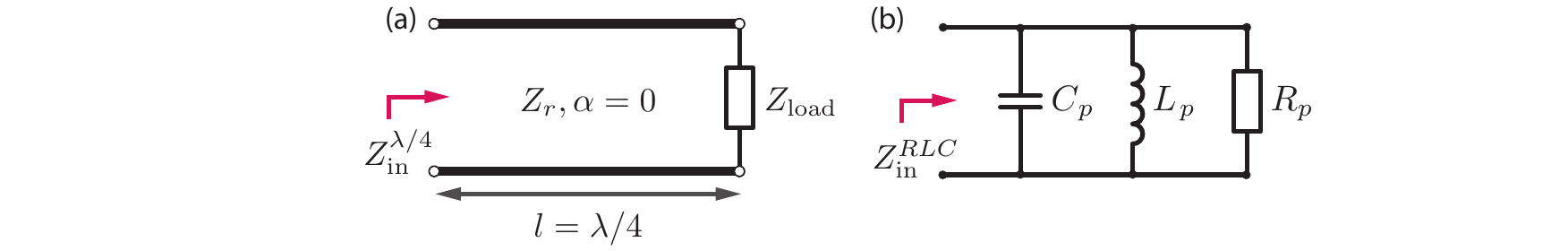}
\caption{
(a) Loaded quarterwave transmission line.  (b) Parallel $RLC$-circuit.}
\label{fig:comparison_circuit}
\end{figure}


\subsection{\label{subsec:loadedquarterres}Loaded $\lambda/4$-resonator}
In the following we consider a $\lambda/4$-resonator, in which the normally shorted end is replaced by a load impedance $Z_{\rm{load}}$ as shown in Fig.~\ref{fig:comparison_circuit}(a). 
In general, the input impedance of a transmission line (TL) of length $l$ and characteristic impedance $Z_r$ shunted by $Z_{\rm{load}}$ is given by
\begin{equation}
\label{eq:inputtl}
Z_{\rm{in,TL}}=Z_r\frac{Z_{\rm{load}}+Z_r\tanh(\gamma{}l)}{Z_r+Z_{\rm{load}}\tanh(\gamma{}l)}.
\end{equation}
Since the complex propagation constant $\gamma$ can be expressed as $\gamma=\alpha+j\beta$, where the real part $\alpha$ is the attenuation constant in TL and, the imaginary part $\beta$ the wavenumber of the TL, we can rewrite
\begin{equation}
\tanh(\gamma{}l)=\frac{1-j\tanh(\alpha{}l)\cot(\beta{}l)}{\tanh(\alpha{}l)-j\cot(\beta{}l)}.
\end{equation}
With $\beta=\omega/v_{\textrm{p}}$, where $v_{\textrm{p}}$ is the phase velocity of the TL and by introducing the relative frequency $\delta\omega=\omega-\omega_0$ with respect to the resonance frequency $\omega_0$, the argument of the $\cot$-term becomes
\begin{equation}
\beta{}l=\frac{\omega_0l}{v_{\textrm{p}}}+\frac{\delta\omega l}{v_{\textrm{p}}}.
\end{equation}
The phase velocity at resonance for a quarterwave resonator (${\lambda=4l}$) reads $v_{\textrm{p}}=\lambda{}f_0=2l\frac{\omega_0}{\pi}$ and therefore
\begin{equation}
\beta{}l=\frac{\pi}{2}+\frac{\pi\delta\omega}{2\omega_0}.
\end{equation}
Assuming $\delta\omega$ being small, we can approximate
\begin{equation}
\cot(\beta{}l)=\cot\left(\frac{\pi}{2}+\frac{\pi\delta\omega}{2\omega_0}\right)=-\tan\left(\frac{\pi\delta\omega}{2\omega_0}\right)\approx-\frac{\pi\delta\omega}{2\omega_0}.
\end{equation}
Assuming that the $\lambda/4$-resonator is lossless ($\alpha=0$) we can simplify the input impedance to
\begin{equation}
Z_{\textrm{in}}^{\lambda/4}=Z_r\frac{Z_{\rm{load}}-jZ_r\frac{2\omega_0}{\pi\delta\omega}}{Z_r-jZ_{\rm{load}}\frac{2\omega_0}{\pi\delta\omega}}.
\end{equation}
In the case of $Z_r\gg\frac{\pi\delta\omega}{2\omega_0}Z_{\rm{load}}$, we can write
\begin{equation}
\label{eq: tlr: inim_4w_approx}
Z_{\textrm{in}}^{\lambda/4}=\frac{1}{\frac{Z_{\rm{load}}}{Z_r^2}+j\frac{\pi\delta\omega}{2\omega_0Z_r}}.
\end{equation}
In general the load impedance is complex valued and can be decomposed into its real and imaginary part, such that 
${Z_{\rm{load}}=\textrm{Re}(Z_{\rm{load}})+j\textrm{Im}(Z_{\rm{load}})}$, which leads to
\begin{equation}
\label{eq: tlr: inim_4w_approx_load}
Z_{\textrm{in}}^{\lambda/4}=\frac{1}{\frac{\textrm{Re}(Z_{\rm{load}})}{Z_r^2}+\frac{j}{Z_r}\left[\frac{\pi\delta\omega}{2\omega_0}+\frac{\textrm{Im}(Z_{\rm{load}})}{Z_r}\right]}.
\end{equation}


\subsection{\label{subsec:inputRLC}Input impedance of parallel $RLC$-circuit}
We compare this now to a parallel $RLC$ resonant circuit shown in Fig.~\ref{fig:comparison_circuit}(b). The input impedance of this circuit simply reads

\begin{equation}
\label{eq: tlr: inim_RLC_full}
Z_{\textrm{in}}^{RLC}=\left(\frac{1}{R_p}+\frac{1}{j\omega{}L_p}+j\omega{}C_p\right)^{-1}
\end{equation}
and its resonance frequency is $\omega_0=1/\sqrt{L_pC_p}$. Making again use of the relative frequency shift and analysing the circuit near resonance allows us to rewrite the former equation to
\begin{equation}
\label{eq: tlr: inim_RLC_approx}
Z_{\textrm{in}}^{RLC}\approx\frac{1}{1/R_p+2j\delta\omega{}C_p}
\end{equation}
or alternatively as
\begin{equation}
\label{eq: tlr: inim_RLC_approx2}
Z_{\textrm{in}}^{RLC}\approx\frac{1}{1/R_p+2j\sqrt{\frac{C_p}{L_p}}\left(\frac{\delta\omega}{\omega_0}\right)}
\end{equation}
Additionally, the internal quality factor of the parallel resonant circuit can be expressed as
\begin{equation}
\label{eq:rlcq}
Q^{RLC}=\omega_0R_pC_p.
\end{equation}


\subsection{\label{subsec:undloadedquarterres}Unloaded $\lambda/4$-resonator}
We can describe the special case of an unloaded quarterwave resonance circuit by evaluating Eq.~\ref{eq: tlr: inim_4w_approx} for $Z_{\rm{load}}=0$:
\begin{equation}
\label{eq: tlr: inim_4w_approx_short}
Z_{\textrm{in}}^{\lambda/4}=\frac{1}{j\frac{\pi}{2Z_r}\left(\frac{\delta\omega}{\omega_0}\right)}.
\end{equation}
By directly comparing Eq.~\ref{eq: tlr: inim_4w_approx_short} with Eq.~\ref{eq: tlr: inim_RLC_approx2} for $R_p\rightarrow\infty$ one finds 
\begin{equation}
\frac{\pi}{2Z_r}=2\sqrt{\frac{C_p}{L_p}}.
\end{equation}
One can now determine the capacitance of the equivalent parallel $RLC$ circuit with $\omega_0=1/\sqrt{L_pC_p}$ as
\begin{equation}
\label{eq:trl:eqvc}
C_p=\frac{\pi}{4\omega_0Z_r}
\end{equation}
and the inductance of the equivalent circuit as
\begin{equation}
\label{eq:trl:eqvl}
L_p=\frac{4Z_r}{\pi\omega_0}.
\end{equation}

In a real experimental scenario the characteristic impedance $Z_r$ is often not known precisely, since besides geometric ingredients -- in particular, the capacitance per unit length $\mathcal{C}_r$ and the self-inductance per unit length $\mathcal{L}_s$ -- there is also a contribution from material properties, which gives rise to the kinetic inductance per unit length $\mathcal{L}_k$. Consequently, the characteristic impedance reads $Z_r=\sqrt{\mathcal{L}_r/{\mathcal{C}_r}}$, where $\mathcal{L}_r=\mathcal{L}_s+\mathcal{L}_k$. Both $\mathcal{C}_r$ and $\mathcal{L}_s$ can be computed with conformal mapping techniques to very high accuracy, whereas $\mathcal{L}_k$ needs to be determined experimentally. $\mathcal{L}_k$ can be measured in a temperature dependence or estimated via the low temperature normal sheet resistance~\cite{S_Zmuidzinas2012a}.
In order to circumvent this inconvenience, we can make use of the wavelength $\lambda$, which in the case of a lossless transmission line is given by
\begin{equation}
\lambda=\frac{2\pi}{\beta}=\frac{2\pi}{\omega_0\sqrt{\mathcal{L}_r\mathcal{C}_r}}.
\end{equation} 
By rearranging this expression at the quarterwave resonance condition and multiplying both sides with $\mathcal{C}_r$, we find
\begin{equation}
\omega_0\mathcal{C}_r=\frac{2\pi}{4l\sqrt{\mathcal{L}_r\mathcal{C}_r}\frac{1}{\mathcal{C}_r}}.
\end{equation}
Substituting $Z_r=\sqrt{{\mathcal{L}_r}/{\mathcal{C}_r}}$ into the previous equation and solve for $Z_r$ leads to 
\begin{equation}
\label{eq: tlr: Zr_geo}
Z_r=\frac{2\pi}{4l\omega_0\mathcal{C}_r}.
\end{equation}
This is now a description for $Z_r$ by just geometrical means ($l$ and $\mathcal{C}_r$) in combination with the resonance frequency $\omega_0$. Here, we do assume an ideal resonator without any coupling to the environment - however those corrections will be small for large coupling quality factors $Q_c$.
Now inserting Eq.~\ref{eq: tlr: Zr_geo} into the expressions for the equivalent circuit (Eq.~\ref{eq:trl:eqvc}~and~\ref{eq:trl:eqvl}), we find in agreement with Ref.~\cite{S_goeppl2008}
\begin{equation}
\label{eq: tlr: C_geo}
C_p=\frac{\mathcal{C}_rl}{2},
\end{equation}
\begin{equation}
\label{eq: tlr: L_geo}
L_p=\frac{2}{l\omega_0^2\mathcal{C}_r}.
\end{equation}


\subsection{\label{subsec:eva_charZ_pInd}Evaluating $Z_r$ and $L_p$}
\begin{figure}[!h]
\includegraphics[width=0.8\columnwidth]{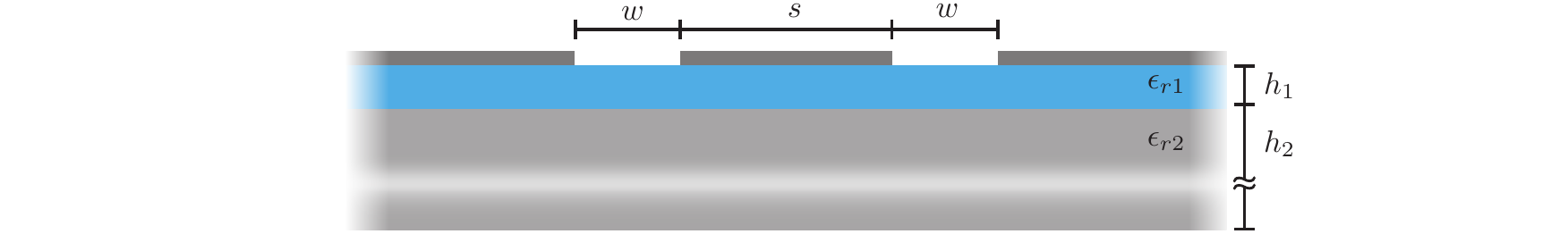}
\caption{\label{fig:cross_sec_tl} Cross-section of a transmission line on a layered substrate.
}
\end{figure}

In order to evaluate characteristic properties of the resonant circuit, we make use of conformal mapping techniques derived in Ref.~\cite{S_Gevorgian1994} to express the capacitance per unit length. 
The effective dielectric constant of a two-layered substrate is found to be
\begin{equation}
\widetilde{\epsilon_{\rm{eff}}}=1+\frac{\epsilon_{r1}-\epsilon_{r2}}{2}\cdot\frac{K(k_1)K(k_0^{\prime})}{K(k_1^{\prime})K(k_0)}+\frac{\epsilon_{r2}-1}{2}\cdot\frac{K(k_2)K(k_0^{\prime})}{K(k_2^{\prime})K(k_0)}
\end{equation}
and the corresponding capacitance per unit length reads
\begin{equation}
\mathcal{C}_r=4\epsilon_0\widetilde{\epsilon_{\rm{eff}}}\frac{K(k_0)}{K(k_0^{\prime})}.
\end{equation}
The functions $K$ are the complete elliptical integrals of the first kind, in which
\begin{align*}
k_0=&\frac{s}{s+2w}\\
k_1=&\frac{\sinh\left(\frac{\pi s}{4h_1}\right)}{\sinh\left(\frac{\pi (s+2w)}{4h_1}\right)}\\
k_2=&\frac{\sinh\left(\frac{\pi s}{4(h_1+h_2)}\right)}{\sinh\left(\frac{\pi (s+2w)}{4(h_1+h_2)}\right)}\\
k_i^{\prime}=&\sqrt{1-k_i^2}~~~~~{\rm{with}}~i=0,1,2,
\end{align*}
where $s$ is the central conductor width, $w$ is the spacing to the ground plane, $h_1$ is the thickness of the top dielectric with relative permittivity $\epsilon_{r1}$ and $h_2$ is the thickness of the bottom dielectric with relative permittivity $\epsilon_{r2}$ see Fig.\,\ref{fig:cross_sec_tl}. 

With $s=12.1\,\mu$m, $w=6.1\,\mu$m, SiO$_2$ thickness $h_1=170$\,nm, Si thickness $h_2=500\,\mu$m, SiO$_2$ permittivity $\epsilon_{r1}=3.9$, Si permittivity $\epsilon_{r2}=11.8$ and the vacuum permititvity $\epsilon_0=8.854\times10^{-12}$\,F/m we find $\mathcal{C}_r=153.9$\,pF/m. With this and the length of the TL $l=7.54$\,mm in combination with the resonance frequency $f_0\approx3.098029$\,GHz we can now evaluate $Z_r=69.54\,\Omega$ with Eq.\,\ref{eq: tlr: Zr_geo}, $C_p=580$\,fF with Eq.\,\ref{eq: tlr: C_geo}  and $L_p=4.548$\,nH with Eq.\,\ref{eq: tlr: L_geo}.
Note that $\widetilde{\epsilon_{\rm{eff}}}$ describes purely the dielectric properties of the TL, whereas $\epsilon_{\rm{eff}}$ also contains properties of the kinetic inductance.

\section{\label{sec:influence_load}Influence of $Z_{\rm{load}}$ on $f_0$ and $Q_{\rm{load}}$}
The resonance condition for a loaded $\lambda/4$-resonator is fulfilled, when $\textrm{Im}[Z_{\textrm{in}}^{\lambda/4}]=0$, which leads to $\frac{\pi\delta\omega}{2\omega_0}+\frac{\textrm{Im}[Z_{\rm{load}}]}{Z_r}=0$ deduced from Eq.\,\ref{eq: tlr: inim_4w_approx_load}. With the load -- the new resonance frequency is called $\omega_0(=2\pi f_0)$, while the resonance frequency of the unloaded resonator is called $\omega_{\rm{bare}}(=2\pi f_{\rm{bare}})$, hence $\delta\omega=\omega_0-\omega_{\rm{bare}}$. In the limit, $f_0\approx f_{\rm{bare}}$, we can express the resonance frequency shift influenced by the load impedance as
\begin{equation}
\label{eq:shiftfull}
\delta f_0=f_0-f_{\rm{bare}}=-\frac{2}{\pi Z_r}\textrm{Im}(Z_{\rm{load}})f_{\rm{bare}}.
\end{equation}
In order to express the quality factor $Q_{\rm{load}}$ of a loaded quarterwave resonator, we assume that $\frac{\textrm{Im}[Z_{\rm{load}}]}{Z_r}\ll1$, such that we obtain from Eq.\,\ref{eq: tlr: inim_4w_approx_load}
\begin{equation}
Z_{\textrm{in}}^{\lambda/4}\approx\frac{1}{\frac{\textrm{Re}(Z_{\rm{load}})}{Z_r^2}+\frac{j}{Z_r}\left(\frac{\pi\delta\omega}{2\omega_0}\right)}.
\end{equation}
By comparing this expression with Eq.~\ref{eq: tlr: inim_RLC_approx2}, we can conclude that $R_p=\frac{Z_r^2}{\textrm{Re}(Z_{\rm{load}})}$. Combining this finding with Eq. \ref{eq:rlcq} and Eq. \ref{eq:trl:eqvc}, we can express the load quality factor as
\begin{equation}
\label{eq:qloadfull}
Q_{\rm{load}}=\frac{\pi Z_r}{4\textrm{Re}(Z_{\rm{load}})}.
\end{equation}


\section{\label{subsec:Zload}Load impedance $Z_{\rm{load}}$}

\begin{figure}[!t]
\includegraphics[width=0.8\columnwidth]{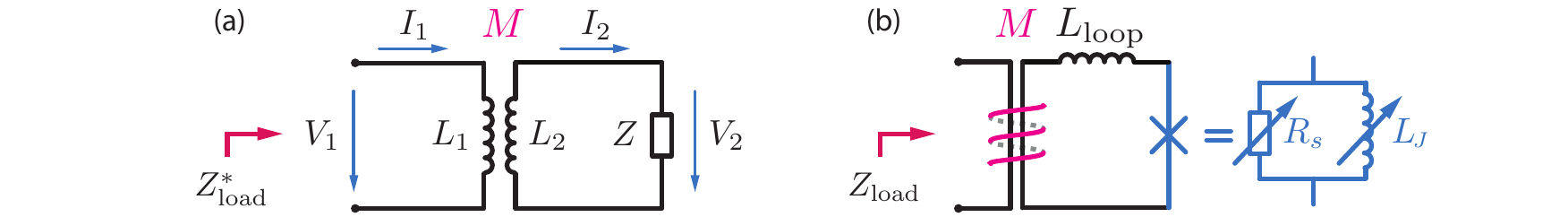}
\caption{
(a) Transformer with the secondary circuit loaded by impedance $Z$. (b) Circuit schematic of the inductively coupled rf SQUID, in which the Josephson junction is modelled as variable Josephson inductance $L_J$ in parallel with a variable shunt resistance $R_s$.
}
\label{fig:load_impedance}
\end{figure}

Until now we have conducted all derivations for a general load impedance terminating the CTL. In the following we derive an expression to describe the impedance provoked by a loop that is inductively coupled to the shorted end of the CTL. First we consider a transformer configuration with a primary part -- the left side in Fig.~\ref{fig:load_impedance}(a) -- that consists of an inductor $L_1$ across which the current $I_1$ flows and the voltage $V_1$ develops.  The secondary part -- the right side in Fig.~\ref{fig:load_impedance}(a) -- describes the mutually coupled loop, which is modelled as an inductance $L_2$ shunted by an impedance $Z$. The mutual inductance $M$ quantifies the coupling strength between the primary and secondary circuit. In the latter the current $I_2$ flows and the voltage $V_2$ appears across $L_2$ and $Z$, respectively.
The voltages in the two circuits can be described by the transformer equations:
\begin{subequations}
\begin{eqnarray}
V_1&=&j\omega{}L_1I_1+j\omega{}MI_2, \label{eq:trans1}
\\
V_2&=&-j\omega{}L_2I_2-j\omega{}MI_1=ZI_2. \label{eq:trans2}
\end{eqnarray}
\end{subequations}
By rearranging the second equality of Eq.~\ref{eq:trans2} to
\begin{equation}
\label{eq:secondarycurrent}
I_2=-\frac{j\omega{}M}{j\omega{}L_2+Z}I_1
\end{equation}
one can express the current in the secondary circuit as a function of the current in the primary. The load impedance seen on the side of the primary or resonator, respectively, can be found by inserting Eq.~\ref{eq:secondarycurrent} into Eq.~\ref{eq:trans1} and divide both sides by $I_1$~\cite{S_terman1955electronic}:
\begin{equation}
\label{eq:mutualzloadfull}
Z^*_{\textrm{load}}=\frac{V_1}{I_1}=j\omega{}L_1+\frac{\omega^2M^2}{j\omega{}L_2+Z}.
\end{equation}
Since the frequency shift is determined by the imaginary part of the load impedance (see Eq.~\ref{eq:shiftfull}) and $L_1$ is constant, the coupling inductance of the primary circuit only provokes an off-set frequency shift. And because the resonance broadening is given by the real part of the load impedance (see Eq.~\ref{eq:qloadfull}), $L_1$ does not influence the load quality factor. Therefore we neglect $L_1$ and absorb its contribution in the bare resonance frequency. With this we obtain a simplified expression for the load impedance: $Z_{\textrm{load}}=\omega^2M^2/\left(j\omega{}L_2+Z\right)$. In the experimental scenario $L_2$ is the inductance of the SQUID loop $L_{\rm{loop}}$ and the shunt impedance $Z$ describes the Josephson junction. Here, we model the junction as a tunable Josephson inductance $L_J$ in parallel with a tunable resistor $R_s$, hence $Z=\left[1/R_s+1/(j\omega{}L_J)\right]^{-1}$. Consequently, the load impedance in our circuit, as shown in Fig.~\ref{fig:load_impedance}(b), reads 
\begin{equation}
\label{eq:mutualzload}
Z_{\textrm{load}}=\frac{\omega^2M^2}{j\omega{}L_{\textrm{loop}}+\left(\frac{1}{R_s}+\frac{1}{j\omega{}L_J}\right)^{-1}}.
\end{equation}
With the load impedance given in Eq.~\ref{eq:mutualzload} substituted into Eq.~\ref{eq:qloadfull}, we obtain for the load quality factor
\begin{equation}
\label{eq:Qloadan}
Q_{\textrm{load}}=\frac{\pi}{4}\cdot\frac{Z_r}{R_s{}M^2}\left[L_{\textrm{loop}}^2+\frac{(L_J+L_{\textrm{loop}})^2{}R_s^2}{\omega^2L_J^2}\right].
\end{equation}
Hence, we found a formalism to convert the load quality factor into an effective lumped element model describing the Josephson junction.
In the case of $L_J\gg{}L_{\textrm{loop}}$ and $\frac{R_s}{\omega{}L_{\textrm{loop}}}\gg{}1$, we can make the approximation:
\begin{equation}
\label{eq:Qloadanapprox}
Q_{\textrm{load}}\approx\frac{\pi}{4}\cdot\frac{Z_r}{\omega^2{}M^2}\cdot{}R_s.
\end{equation}
Assuming $R_s\rightarrow \infty$ in Eq.\,\ref{eq:mutualzload} and making use of Eq.\,\ref{eq:trl:eqvl}, we can approximate Eq.\,\ref{eq:shiftfull} as
\begin{equation}
\label{eq:fresan}
\delta f_0\approx\frac{8}{\pi^2}\frac{M^2}{L_p(L_J+L_{\textrm{loop}})}f_{\textrm{bare}},
\end{equation}
which describes the frequency shift as a function of the Josephson inductance $L_J$, which, in turn, is directly related to the current-phase relation (CPR). In general, this last approximation is not need, but reduces the computational effort tremendously, especially when the iterative screening correction procedure is conducted. 

\section{\label{subsec:test_ana_formula}Compare the analytical expressions with numerical results}
In the following we prove the validity of the analytic formulas for $\delta f_0$ and $Q_{\textrm{load}}$ (Eq.\,\ref{eq:fresan} and Eq.\,\ref{eq:Qloadan}) by comparing their solutions with the numerically evaluated full model. In particular, we generate reflection curve maps and extract from those, the resonance frequency $f_0^{\textrm{full}}$ and the load quality factor $Q_{\textrm{load}}^{\textrm{full}}$ by fitting as explained in Sec.\,\ref{sec:rescurvefit}.
In general, the reflection coefficient reads
\begin{equation}
\label{eq:gammafull}
\Gamma=\frac{Z_{\textrm{in}}^{\textrm{full}}-Z_0}{Z_{\textrm{in}}^{\textrm{full}}+Z_0},
\end{equation} 
where $Z_0=50\,\Omega$ is the environmental impedance. 
The coupling capacitance $C_c$ between the measurement set-up and the TL leads to an impedance $Z_c=1/(j2\pi f C_c)$ in series with the input impedance of the loaded TL, $Z_{\rm{in,TL}}$, such that 
\begin{equation}
\label{eq:couplingcin}
Z_{\textrm{in}}^{\textrm{full}}=Z_c+Z_{\rm{in,TL}}.
\end{equation}
Hence, by combining Eq.\,\ref{eq:inputtl} and Eq.\,\ref{eq:mutualzload} in Eq.\,\ref{eq:gammafull} with the use of Eq.\,\ref{eq:couplingcin}, we can express $\Gamma$ as a function of $Z_{\textrm{load}}$ with properties of the TL.

\begin{figure}[!h]
\includegraphics[width=1\columnwidth]{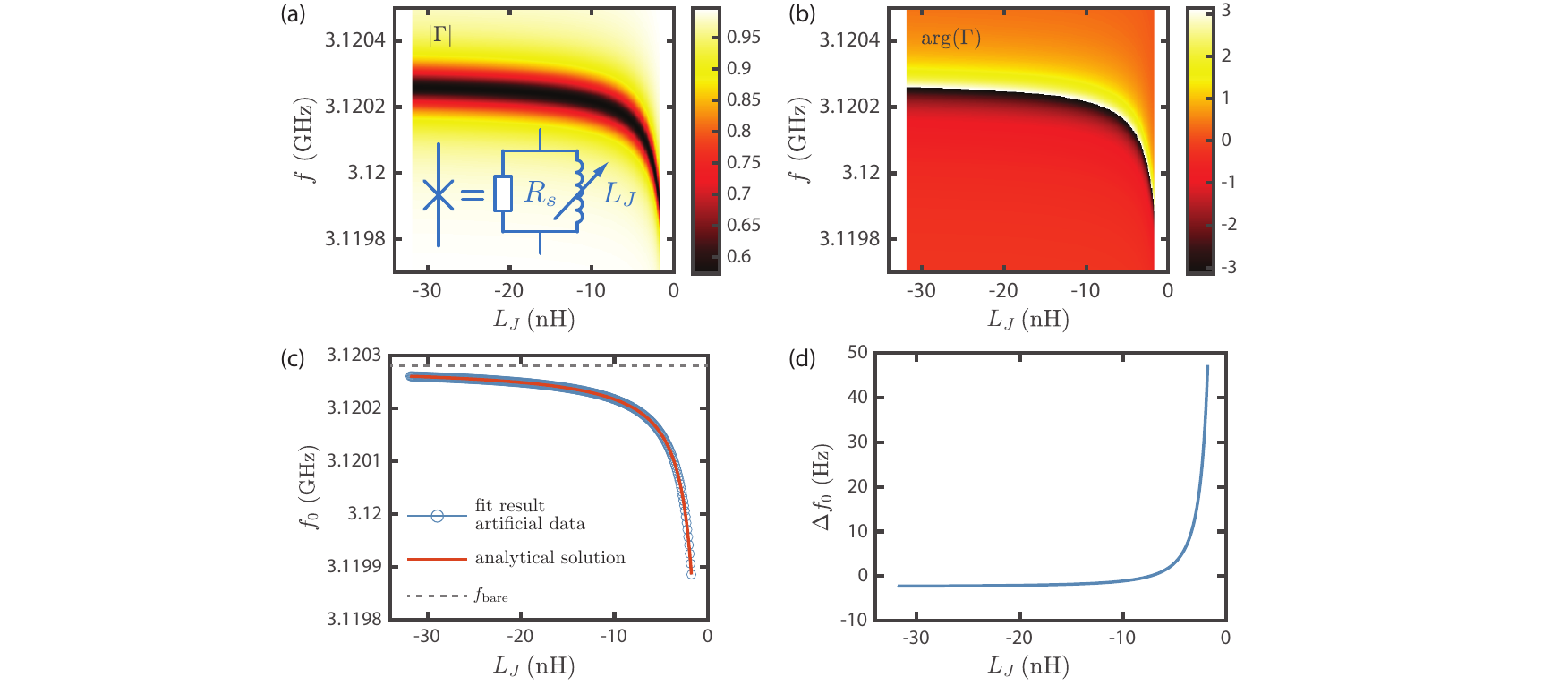}
\caption{Parameters for artificial data: $C_c=4.6$\,fF, $\alpha=0.001$\,m$^{-1}$, $\beta=2\pi f\frac{\sqrt{\epsilon_{\rm{eff}}}}{c}$, where ${\epsilon_{\rm{eff}}=11.225}$ and $c$ is the speed of light, $l=7.1$\,mm, $Z_r=64.5\,\Omega$, $L_{\rm{loop}}=200$\,pH, \hbox{$M=32$\,pH}, \hbox{$R_s=100$\,M$\Omega$}, sinusoidal CPR $\rightarrow L_J=\frac{2\pi}{\Phi_0 I_c\cos(\varphi)}$, here the sweep range corresponds to the phase biasing condition $\varphi=\pi$ and the critical current is tuned $I_c=10\rightarrow 180$\,nA (larger $I_c$ produces more shift).
(a)-(b) Colormaps of the artificial data $|\Gamma|$ and $\arg(\Gamma)$ as a function of Josephson inductance $L_J$.
(c) The resonance frequency $f_0^{\textrm{full}}$ (blue circles) obtained by fitting the artificial resonance curves. The analytically predicted resonance frequency (red lines) deduced from Eq.\,\ref{eq:fresan} with the same parameters as listed above and $f_{\rm{bare}}=3.12028$\,GHz obtained from minimizing $|\Gamma|$ for $L_J\rightarrow\infty$ in the full model.
(d) Difference between $f_0^{\textrm{full}}$ and the analytically obtained resonance frequency.}
\label{fig:Ana_sim_fres}
\end{figure}

First, we provide a consistency proof for the expression of the frequency shift. 
In Fig.~\ref{fig:Ana_sim_fres}, we keep the shunt resistance $R_s=100$\,M$\Omega$ constant and sweep the Josephson inductance $L_J$. 
From the artificial $\Gamma$-maps shown in Fig.~\ref{fig:Ana_sim_fres}(a)-(b) a clear change in $f_0^{\textrm{full}}$ is observed as a function of $L_J$, while the lineshape is not affected. Details about the parameters used here are listed in the figure caption. 
In Fig.~\ref{fig:Ana_sim_fres}(c) we overlay the fit results for resonance frequency $f_0^{\textrm{full}}$ of the artificial data (blue circles) with the prediction from the analytic formalism (red, Eq.\,\ref{eq:fresan}). From Fig.~\ref{fig:Ana_sim_fres}(d), which shows the difference $\Delta f_0$ between the resonance frequency of the artificial data and the one obtained from the analytic formalism, we observe only slight discrepancies on the order of Hz. Consequently, Eq.\,\ref{eq:fresan} describes the resonance frequency as function of $L_J$ to a very high accuracy.

\begin{figure}[!t]
\includegraphics[width=1\columnwidth]{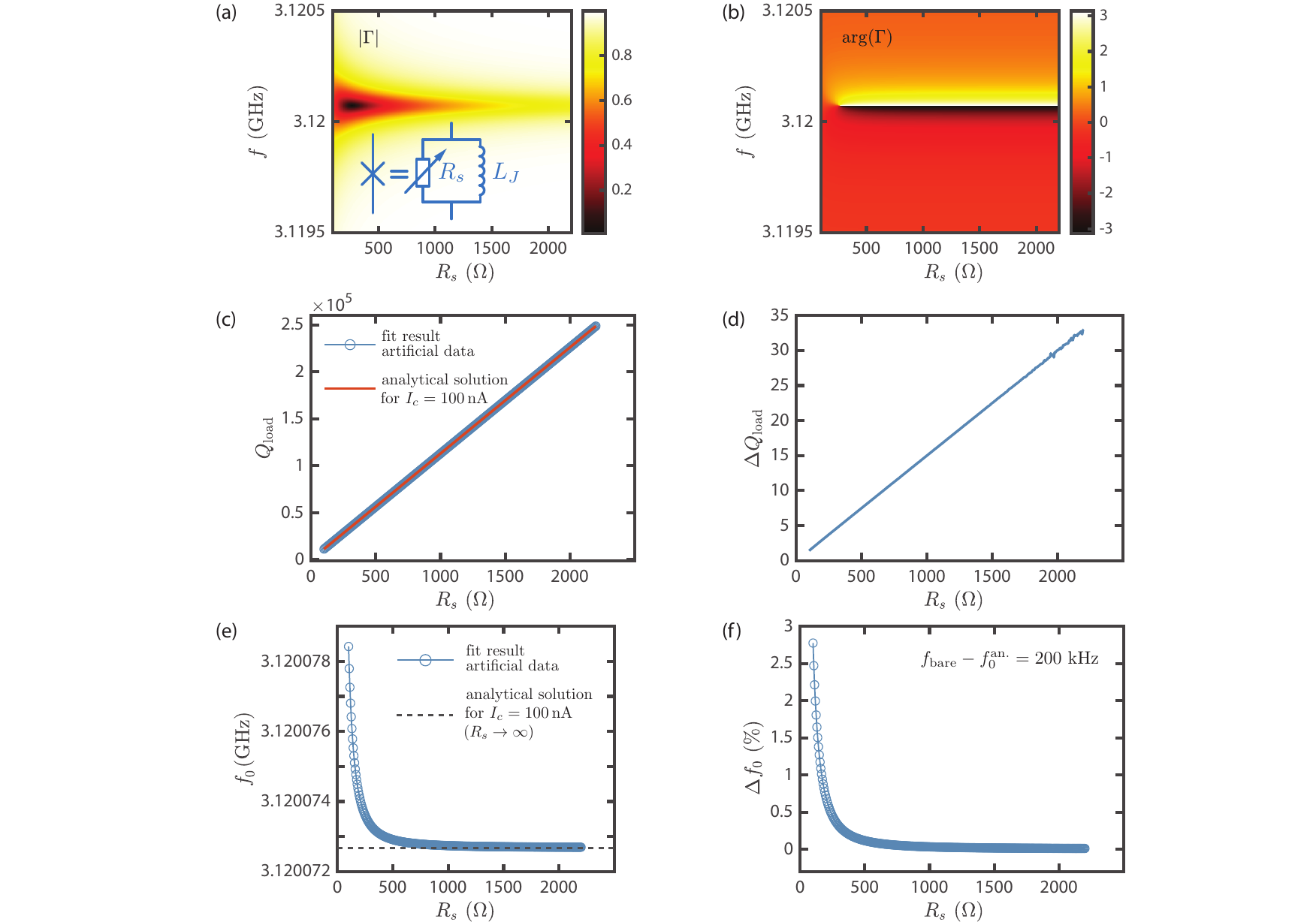}

\caption{Parameters for artificial data: $C_c=4.6$ fF, $\alpha=0$, $\beta=2\pi f\frac{\sqrt{\epsilon_{\rm{eff}}}}{c}$, where $\epsilon_{\rm{eff}}=11.225$ and $c$ is the speed of light, $l=7.1$ mm, $Z_r=64.5~\Omega$, $L_{\rm{loop}}=200$ pH, $M=32$ pH, $I_c=100$ nA, $L_J=-\Phi_0/(2\pi I_c)=-3.183$ nH.
(a)-(b) Colormaps of the artificial data $|\Gamma|$ and $\arg(\Gamma)$ as a function of shunt resistance $R_s$.
(c) The load quality factor $Q_{\textrm{load}}^{\textrm{full}}$ (blue circles) obtained by fitting the artificial resonance curves as a function of $R_s$. The analytically predicted load quality factor (red lines) deduced from Eq.\,\ref{eq:Qloadan} with the same parameters as listed above and $f_{\rm{bare}}=3.12028$\,GHz obtained from minimizing $|\Gamma|$ for $L_J\rightarrow\infty$ and $R_s\rightarrow\infty$ in the full model.
(d) Difference between $Q_{\textrm{load}}^{\textrm{full}}$ and the analytically obtained load quality factor.
(e) The resonance frequency $f_0^{\textrm{full}}$ (blue circles) obtained by fitting the artificial resonance curves and predicted resonance frequency for $L_J=-3.2$\,nH.
(f) Relative error between the actual resonance frequency shift and the predicted resonance frequency shift $f_0^{\textrm{an.}}$; $\Delta f_0(\%)=(f_0^{\textrm{full}}-f_0^{\textrm{an.}})/({f_{\rm{bare}}-f_0^{\textrm{an.}}})$. 
}
\label{fig:Ana_sim_Qload}
\end{figure}

Second, we provide a consistency proof for the expression of the load quality factor.
In Fig.~\ref{fig:Ana_sim_Qload}, we keep the Josephson inductance $L_J=-3.2$\,nH (sinusoidal CPR with $I_c=100$\,nA at $\varphi=\pi$) constant and sweep the shunt resistance $R_s$. For simplicity we set $\alpha=0$, such that the effective quality factor is determined by the load. 
From the artificial $\Gamma$-maps shown in Fig.~\ref{fig:Ana_sim_Qload}(a)-(b) a clear change the lineshape of the resonance curve as a function of $R_s$ is observed. Details about the parameters used here are listed in the figure caption. 
The dark region in Fig.~\ref{fig:Ana_sim_Qload}(a), 
where $|\Gamma|=0$ corresponds to full matching, where $Q_c=Q_{\textrm{load}}$. For small $R_s$ values the resonator becomes overcoupled ($Q_c>Q_{\textrm{load}}$) and $\arg(\Gamma)$ evolves smoothly, whereas for large $R_s$ values the resonator becomes undercoupled ($Q_c<Q_{\textrm{load}}$) and $\arg(\Gamma)$ undergoes a $2\pi$-leap.  The coupling quality factor can be expressed as $Q_c=\frac{\pi}{4\omega^2Z_0Z_rC_c^2}$,
for which we find $Q_c=29\,950$ with the model parameters $C_c=4.6$\,fF, $Z_r=64.5\,\Omega$ and $\omega\approx2\pi\cdot3.12$\,GHz. 
In Fig.~\ref{fig:Ana_sim_Qload}(c) we overlay the fit results for $Q_{\textrm{load}}$ of the artificial data (blue circles) and the prediction from the analytic formalism (red, Eq.~\ref{eq:Qloadan}). From Fig.~\ref{fig:Ana_sim_Qload}(d), which presents the difference $\Delta Q_{\textrm{load}}$ between the artificial data and the predications, we observe very small discrepancies. 
Since, $R_s$ is naturally present in $\textrm{Im}(Z_{\rm{load}})$, changing the resistance causes in addition a small shift of the resonance frequency. 
By the comparison between the resonance frequency of the artificial data and the one obtained analytically (assumption $R_s\rightarrow\infty$, such that Eq.~\ref{eq:fresan} becomes valid) shown in Fig.~\ref{fig:Ana_sim_Qload}(e), we observe a discrepancy of $\sim6$\,kHz for the smallest $R_s$ value. 
On a first glance this seems a lot, one should however relate this number with the overall shift of the resonance frequency coming from $L_J=-3.2$\,nH, which is about $200$\,kHz. Hence, the error induced by neglecting $R_s$, is on the order of a few $\%$ as long as $R_s\geq100\,\Omega$, which is the case for our measurement.

\clearpage


\section{\label{sec:screening_correction}Curve fitting with screening correction}

Using Eqs. \ref*{eq:fres}, \ref*{eq:LCR} and \ref*{eq:IsFourier} from the maintext, we express the shifted resonance frequency $f_0$ as a function of the junction phase $\varphi$,

\begin{equation}
\label{eq:fres_vs_phi}
f_0 (\varphi) = \left[  \frac{8}{\pi^2}\frac{M^2}{L_p\left( \left( \frac{2\pi}{\Phi_0} \sum_{k=1}^{k_{\rm{max}}} (-1)^{k-1}A_{k} k \cos(k\varphi) \right) ^{-1} 
+L_{\rm{loop}} \right)} +1 \right] f_{\rm{bare}}.
\end{equation}
We fix the values of $M=30.83$ pH, $L_{\rm{loop}}=211$ pH and $L_p=4.546$ nH obtained from simulations, and treat $A_k$ and $f_{\rm{bare}}$ as free fitting parameters.

In the absence of current in the rf SQUID loop, the junction phase $\varphi$ is solely determined by the external magnetic flux $\Phi$ in the loop, $\varphi = \varphi_{\rm{ext}}=2\pi\Phi/\Phi_0$. Taking into account the flux created by the circulating DC supercurrent yields

\begin{equation}\label{eq:screening_SI}
\varphi=\varphi_{\rm{ext}}-\frac{2\pi}{\Phi_0}L_{\rm{loop}}I_s(\varphi).
\end{equation} 
This means, that the junction phase $\varphi$ depends on the external flux and the CPR to-be-determined as well.

In the experiment the resonance frequency $f_0$ is measured as a function of the current in the flux line. Using the periodicity of the signal, we convert the flux current to the external phase $\varphi_{\rm{ext}}$ by applying a linear transformation. Next, to determine the CPR from the $(f_0, \varphi_{\rm{ext}})$ data while taking into account the flux contribution of the supercurrent, we find the self-consistent solution of Eqs.\,\ref{eq:fres_vs_phi} and \ref{eq:screening_SI} with an iterative method. The scheme is presented with the pseudocode in Algorithm~\ref{algo:screening}. Essentially, it combines fixed-point iteration with Eq.\,\ref{eq:screening_SI} and least-square fits to Eq.\,\ref{eq:fres_vs_phi}. The procedure realizes the non-linear transformation of $\varphi_{\rm{ext}}$  to $\varphi$, and outputs the harmonic coefficients $A_k$ and the bare resonance frequency $f_{\rm{bare}}$.

\begin{algorithm}[H]
	\caption{Iterative procedure for curve fitting with screening correction}\label{alg:qvol}
	\begin{algorithmic}
		\Function{fitWithScreening}{$ f_0, \varphi_{\rm{ext}} ; M, L_p, L_{\rm{loop}}, n_{\rm{iter}} = 30, \alpha = 0.2 \ldots 0.4, {k_{\rm{max}}}=10$}
		\State $\varphi = \varphi_{\rm{ext}} $ \Comment{Initialization}
 		\For{$n_{\rm{iter}}$ repetitions}
		\State $A_k, f_{\rm{bare}} \gets$ least-square fit of $(f_0,\varphi)$ data points to Eq.(\ref{eq:fres_vs_phi})
		\State $I_s (\varphi) = \sum_{k=1}^{k_{\rm{max}}} (-1)^{k-1}A_{k} \sin(k\varphi) $ \Comment{Substitution of $A_k$ into Eq. \ref*{eq:IsFourier}}
		\State $\varphi^{\rm{new}} = \varphi_{\rm{ext}}-{2\pi}/{\Phi_0} \cdot L_{\rm{loop}}I_s(\varphi) $ \Comment{Substitution of $I_s$ into Eq. \ref{eq:screening_SI}}
		\State $\varphi = \alpha \varphi^{\rm{new}} + (1-\alpha) \varphi $ \Comment{Smooth update} 
		\EndFor
		\State \Return $A_k, f_{\rm{bare}}, \varphi$
		\EndFunction
	\end{algorithmic}
	\label{algo:screening}
\end{algorithm}

\begin{figure}[!b]
\includegraphics[width=0.9\columnwidth]{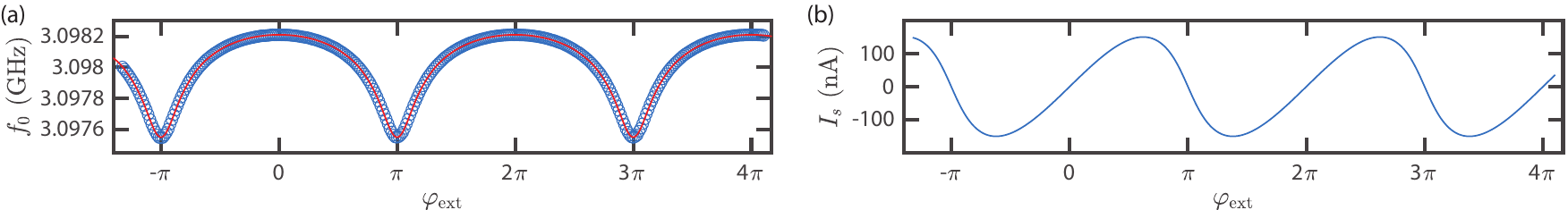}
\caption{(a) Extracted resonance frequency $f_0$ (blue circles) as a function of external phase $\varphi_{\rm {ext}}$ at $V_{\rm{bg}}=6$\,V. From the fit (solid red, Eq.\,\ref{eq:fres_vs_phi}) one obtains the supercurrent $I_s(\varphi_{\rm {ext}})$ as a function of external phase shown in (b).
}
\label{fig:screening_fit_fres}
\end{figure}

\begin{figure}[!h]
\includegraphics[width=0.8\columnwidth]{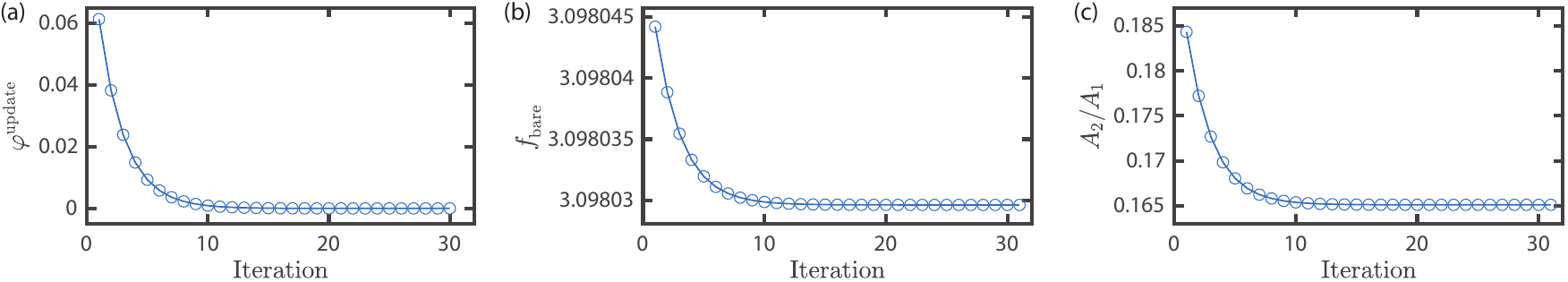}
\caption{
Convergence of the iterative curve fitting method ($V_g = 6$~V, $\alpha=0.4$). (a) The magnitude of the phase update, $\varphi^{\rm{update}} = \langle |\varphi^{\rm{new}} -\varphi | \rangle _{\rm avg}$ converges to zero as the iteration progresses. (b-c) Convergence of the bare resonance frequency $f_{\rm{bare}}$ and the harmonic coefficient ratio $A_2/A_1$.}
\label{fig:screening_fit_iteration}
\end{figure}

\begin{figure}[!h]
\includegraphics[width=0.8\columnwidth]{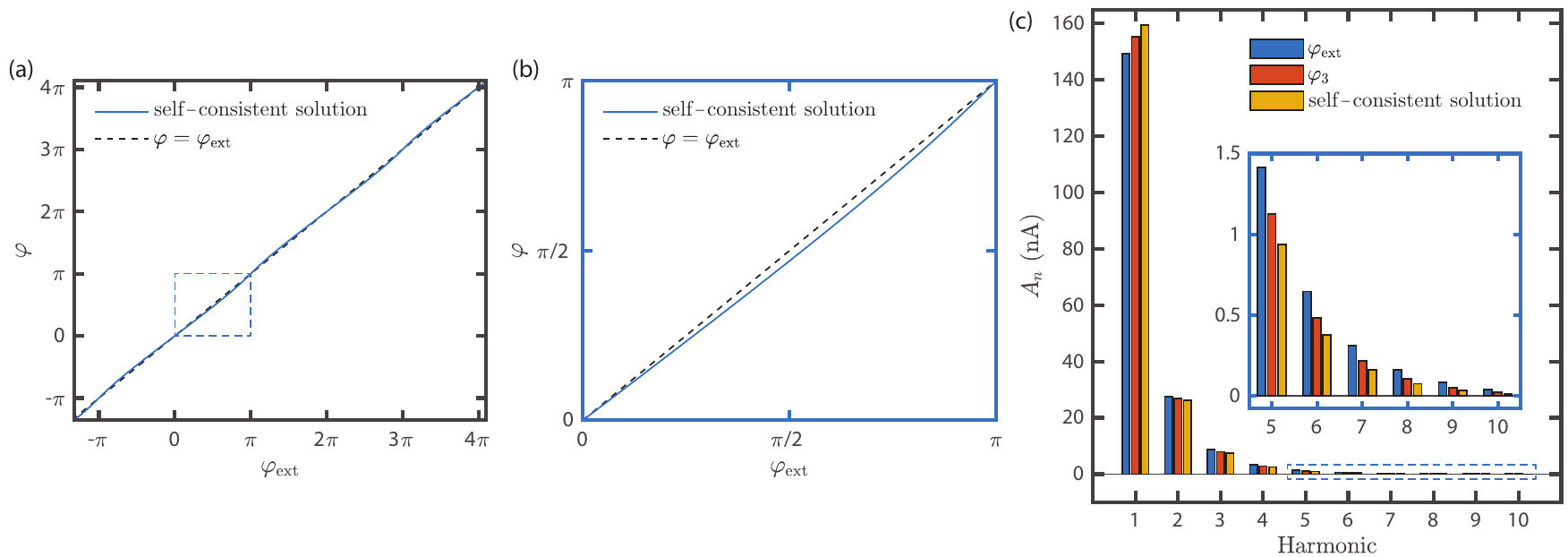}
\caption{
Outcome of the iterative curve fitting method ($V_{\rm{bg}}=6$\,V) (a-b) Junction phase $\varphi$ as a function of the external phase $\varphi_{\rm{ext}}=2\pi\Phi/\Phi_0$. The self-consistent solution of the equation set (continuous, blue) deviates from the $\varphi = \varphi_{\rm{ext}}$ line (dashed black). (c) Harmonic coefficients $A_k$ at different stages of the iteration: initial solution (corresponding to the $\varphi=\varphi_{\rm{ext}}$ approximation), solution at $n_{\rm{iter}} = 3$  and $n_{\rm{iter}} = 30$ (converged).
}
\label{fig:screening_fit_solution}
\end{figure}

In the following we illustrate the fitting routine with the experimental data obtained at $V_{\rm{bg}}=6$\,V. In Fig.\,\ref{fig:screening_fit_fres} the initialization is shown, whereas the iteration and the outcome of the algorithm is illustrated in Figs.\,\ref{fig:screening_fit_iteration}-\ref{fig:screening_fit_solution}. The convergence of the procedure has been checked manually for each gate voltage. Depending on the values of $M$ and $L_{\rm{loop}}$, manual tuning of the smoothing parameter $\alpha$ was necessary.

Figs. \ref{fig:screening_fit_iteration}(c) and \ref{fig:screening_fit_solution}(c) show that neglecting the flux contribution of the supercurrent, and using the approximation $\varphi=\varphi_{\rm{ext}}$ leads to overestimating the skewness of the CPR. While the apparent skewness parameter in this approximation is $S_{\textrm{ext}}=0.2434$, the self-consistent solution yields $S=0.2168$. Similarly, the harmonic coefficient ratio reduces from $A_2/A_1 \approx 0.185$ to $A_2/A_1  \approx 0.165$ as the iteration converges.


\section{\label{sec:chargedensity}Charge carrier density:}
\begin{figure}[!h]
\includegraphics[width=1\columnwidth]{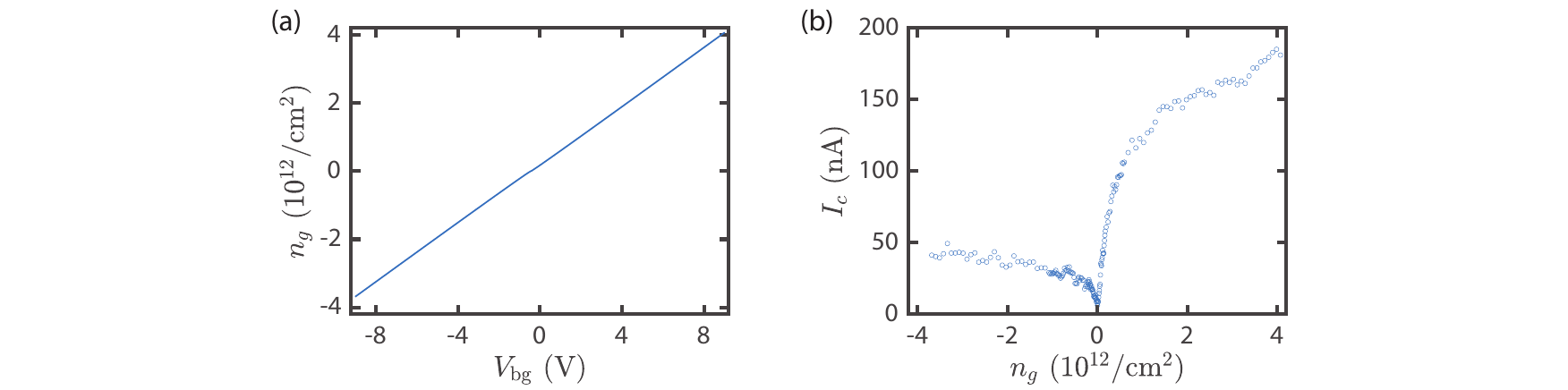}
\caption{(a) Charge carrier density $n_g$ determined with Eq.\,\ref{eq:vtocharge} as function of gate voltage $V_{\rm{bg}}$. (b) Critical current $I_c$ as function of $n_g$.
}
\label{fig:charge_density}
\end{figure}

To convert the applied back gate voltage ($V_{\rm{bg}}$) to charge carrier density ($n_g$) we used a plate capacitor model including the quantum capacitance of graphene \cite{S_Xia2009}, which results in
\begin{equation}
\label{eq:vtocharge}
e(V_{\textrm{bg}}+V_{\textrm{off}})=\frac{e^2n_gd}{\epsilon_0\epsilon_r}+\textrm{sgn}(n_g)\hbar v_F\sqrt{\pi |n_g|},
\end{equation}
where $V_{\rm{off}}=0.44$\,V is the offset voltage of the charge neutrality point with respect to 0\,V, $e$ is the electron charge, $d=47.5$\,nm the thickness of the gate dielectric, $\epsilon_0=8.854\times10^{-12}$\,F/m the vacuum permittivity, $\epsilon_r=3.8$ the dielectric constant of hBN 
\cite{S_Laturia2018}, $\hbar$ the reduced Planck constant, and $v_F=10^6$\,m/s the Fermi velocity of graphene. The quantum capacitance corresponding to the second term on the right hand side of Eq.\,\ref{eq:vtocharge} leads to minor deviations of the linear behavior on $n_g$ with respect to $V_{\textrm{bg}}$ around charge neutrality, as shown in Fig.\,\ref{fig:charge_density}\,(a). By using Eq.\,\ref{eq:vtocharge} the previously extracted critical current $I_c(V_{\textrm{bg}})$ is plotted as a function of $n_g$ in Fig.\,\ref{fig:charge_density}\,(b).  In previous works oscillations of $I_c(n_g)$ were observed for negative densities for high mobility and ballistic graphene Josephson junctions 
\cite{S_Allen2017, S_Calado2015}. 
They arise due to quantum interference of the electrons moving in a Fabry-P\'erot cavity 
\cite{S_Young2009}
, which is formed by potential steps in the graphene. Namely, the graphene is $n^\prime$-doped with electrons close to the contacts given by the work function mismatch of the graphene and the Al boundary, while the bulk of graphene is $p$-doped with holes due to the negative applied $V_{bg}$.  The oscillations show their m$^{th}$ maxima at $\sqrt{n_g}=m\sqrt{\pi}/L$, where $L$ corresponds to the length of the cavity. Nevertheless, no such oscillations were observed in our measurement of $I_c(n_g)$, which indicates that the electron transport is diffusive in our sample.

\section{\label{sec:temp_dep}Temperature dependence:}

\subsection{Theoretical description}

We numerical solve the time-dependent Usadel equation~\cite{S_usadel1970generalized,S_Virtanen2011}, from which we infer the inductive $B_J$ and dissipative $G_s$ microwave response of a short diffusive Josephson junction. 
The theoretical predictions are based on characteristic energy scales: The electronic temperature $T$, the photonic energy  $hf$ irradiating on the junction, the relaxation rate $\gamma$, the superconducting gap $\Delta$ and the Thouless energy $E_T$.

\begin{figure}[!h]
\includegraphics[width=0.8\columnwidth]{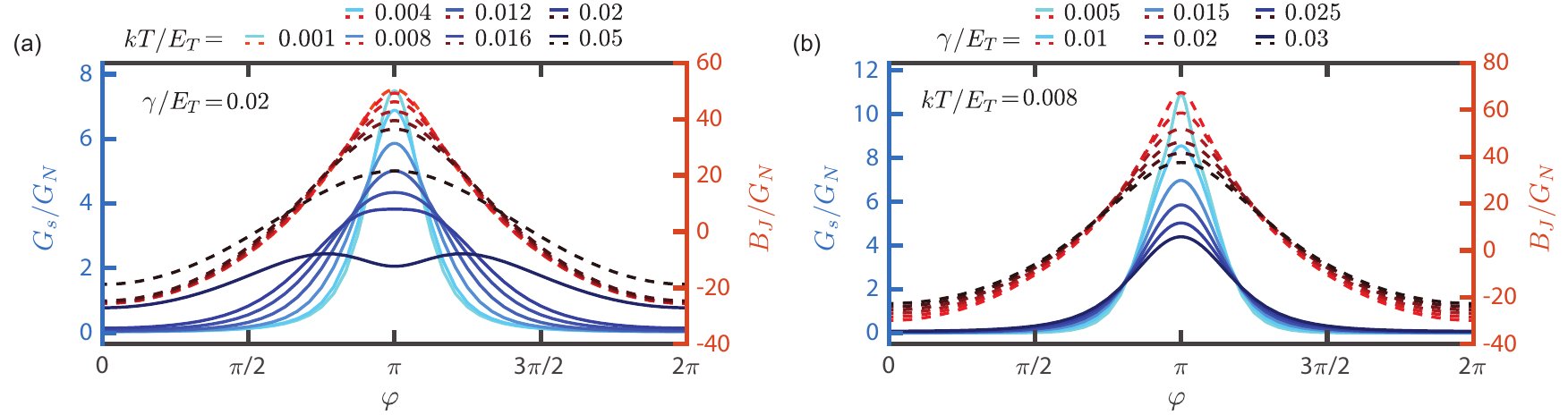}
\caption{
Numerical simulations of the shunt conductance $G_s/G_N$ and junction susceptance $B_J/G_N$ both normalized with the normal state conductance $G_N$. Fixed parameters $\Delta/E_T=0.1$ and $hf/E_T=0.01$. (a) Dissipative (left) and inductive (right) microwave response for different temperatures but fixed relaxation rate. (b) Dissipative (left) and inductive (right) microwave response for different relaxation rates but fixed temperature.
}
\label{fig:temp_sim}
\end{figure}

In Fig.\,\ref{fig:temp_sim}, we fix $\Delta/E_T=0.1$ and $hf/E_T=0.01$. On the left axis the dissipative response normalized with the normal state conductance $G_s/G_N$ is plotted (solid blue lines) and on the right axis the inductive response normalized with the normal state conductance $B_J/G_N$ is plotted (dashed red lines).

In Fig.\,\ref{fig:temp_sim}(a) we fix $\gamma/E_T=0.02$, while we sweep the temperature ratio $kT/E_T$. The wide onset of the dissipation peak even at low temperatures is mainly due to the non-vanishing relaxation ratio $\gamma/E_T$ causing lifetime broadening of the ABS spectrum. With increasing temperature the conductance peak shrinks and becomes wider. A plateau like feature turning into a double wall can be recognized at $\varphi=\pi$ due to the dynamics of the thermally populated $E_n^+$-states.
From the susceptance we observe that the conditions for $B_J/G_N=0$ are moving away from $\varphi=\pi$ for increasing temperature and the absolute values of $B_J/G_N$ at $\varphi=\pi$ and $\varphi=0,2\pi$ approach each other, which means that the CPR is becoming more sinusoidal.

For comparison we present in Fig.\,\ref{fig:temp_sim}(b) the numerical results for fixing $kT/E_T=0.008$, while sweeping the relaxation ratio $\gamma/E_T$. We recognize a less evident change of the dissipation peak center as compared to the temperature sweep. Overall the conductance peak broadens accompanied with a shrinking of the height. Importantly, here the susceptance reveals as well a reduction of the CPR skewness. Note, that this plot is the same as Fig.\,\ref*{fig:spectrum}(d) in the maintext, but globally normalized with $G_N$. Since the temperature effect seem to evolve differently from the relaxation rate dependence -- both of this parameters should be accessible by comparing theoretical predications with experimental data.


\subsection{Experimental results}
From theory it is predicted that for increasing the temperature $T$ the current-phase relation (CPR) becomes more and more sinusoidal, which is due to the balancing between $E_n^+$- and $E_n^-$-states described by the Fermi-Dirac distribution. The population of $E_n^+$-states further affects the absorbency of the ABS spectrum, because transition from $E_n^-\rightarrow E_n^+$ are prohibited if the final state is already occupied. As a consequence the dissipation peak becomes smaller. 

\begin{figure}[!h]
\includegraphics[width=0.8\columnwidth]{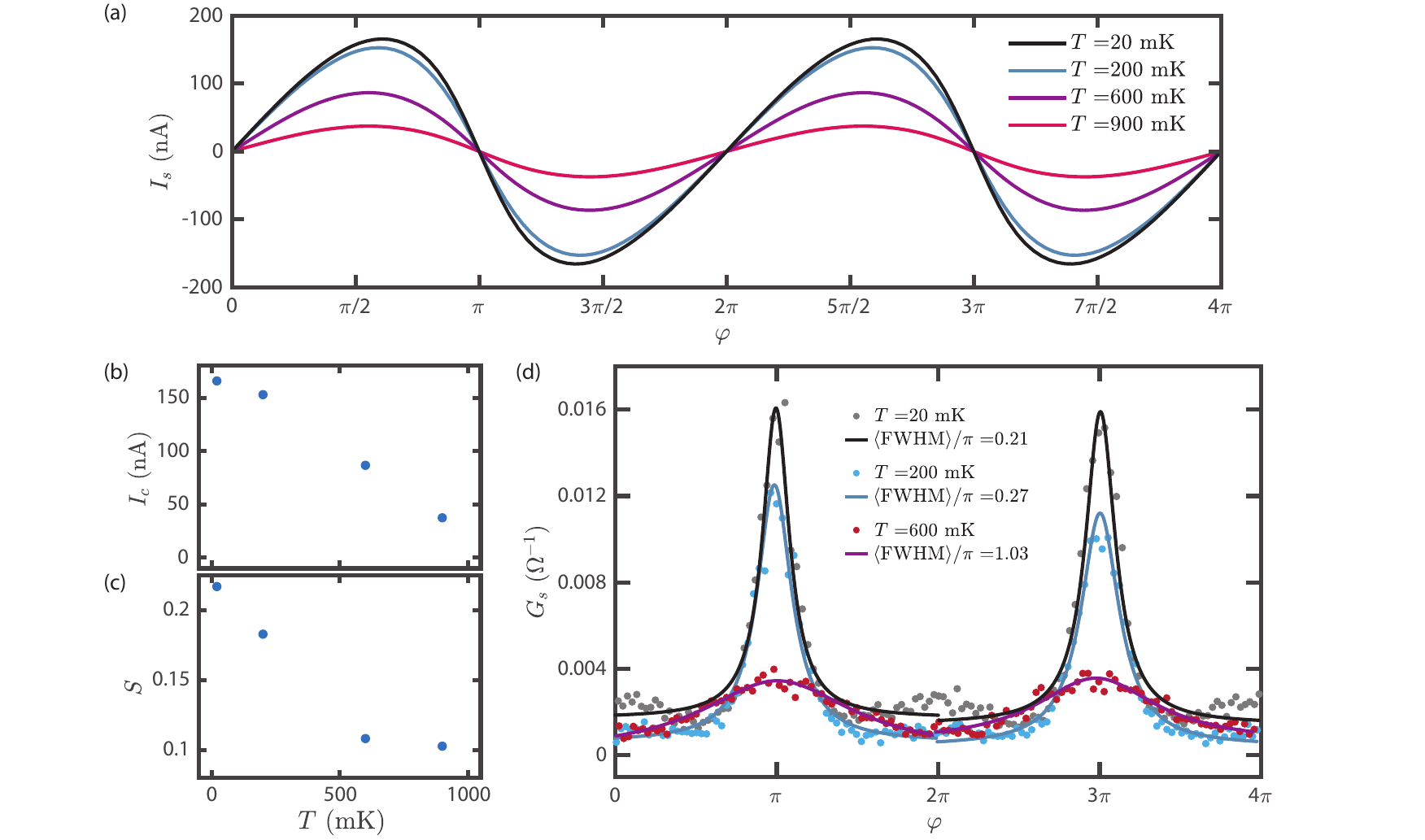}
\caption{Temperature $T$ dependence at $V_{\textrm{bg}}=12$\,V. (a) Self-consistent CPR for different temperatures (see legend). (b) Critical current $I_c$ as a function of $T$. (c) Skewness parameter $S$ as a function of $T$. (d) Phase-dependent dissipation at different $T$. The peaks are fitted with Lorentzian functions, which reveals a clear spreading of the width for increasing $T$.  
}
\label{fig:temp_cpr_lorenz}
\end{figure}

In the following we probe the microwave response of the graphene JJ in terms of the CPR and the phase-dependent dissipation at $V_{\rm{bg}}=12$\,V for different temperatures. We deduce the CPR and the shunt conductance with the same methods described above and in the maintext. 
In Fig.\,\ref{fig:temp_cpr_lorenz}(a) we illustrate the self-consistent CPR solution for different base temperatures adjusted by heating the mixing chamber plate. We observe  
a clearly skewed CPR for temperatures far below the critical temperature of Al ($T_c\approx1.2$\,K), while for $T\rightarrow T_c$ the skewness as well as the critical current $I_c$ decreases as present separately in Figs.\,\ref{fig:temp_cpr_lorenz}(b)-(c). These effects are attributed to: i) the washing out of the energetically low lying states (close to $E=0$), which are responsible for the skewness due to their high transparency and ii) the closing of the superconducting gap. Measurements for $T>900$\,mK were suffering from strong temperature fluctuations.

As postulated by theory we obtain a counter-intuitive decreasing of the dissipation peaks for increasing temperatures as seen in Fig.\,\ref{fig:temp_cpr_lorenz}(d). Not only the height is influenced by the temperature, but also the width, which is also a result from the theoretical predictions. 
We fit the different dissipation peaks with a Lorentzian function of the form $L=\frac{a\left(b/2\right)^2}{(\varphi-c)^2+\left(b/2\right)^2}+d$, where $a$ is a scaler for the peak height, $b$ is the full-width-half-maximum (FWHM), $c$ is a translation on the phase-axis and $d$ describes a vertical offset. We find that the averaged FWHM of the two peaks measured at the same temperature is increasing $\langle {\rm{FWHM}} \rangle\approx 0.2\pi\rightarrow\pi$ for temperatures $T=20\rightarrow600$\,mK, while the peak height shrinks by a factor of $\sim3$.

In contrast to the low temperature results presented in the maintext (Fig.\,\ref*{fig:comp}) we did not find combinations of $kT/E_T$ and $\gamma/E_T$, which simultaneously reproduce the inductive and dissipative response. We attribute this to the granularity of parameter space used in the simulation. In order to describe the microwave behavior of the JJ at high temperatures -- effects like highly enhanced relaxation rates, modifications in junction length limit and the gap closing would need to be considered.

\end{document}